\documentclass[journal=mamobx,manuscript=article]{achemso}
\usepackage[version=3]{mhchem} % Formula subscripts using \ce{}
\usepackage[T1]{fontenc}       % Use modern font encodings
\usepackage{siunitx}
\usepackage{epstopdf}
\usepackage{amssymb}
\setkeys{acs}{usetitle = true}

\usepackage{booktabs}
\renewcommand{\arraystretch}{1.5}
\usepackage[table]{xcolor} 
\usepackage{makecell}

\usepackage{multirow}

\graphicspath{{graphics/}}
\newcommand*{\citen}[1]{%
  \begingroup
    \romannumeral-`\x % remove space at the beginning of \setcitestyle
    \setcitestyle{numbers}%
    \cite{#1}%
  \endgroup   
}

\author{Maud Formanek}
\affiliation{Centro de F\'{i}sica de Materiales (CSIC, UPV/EHU) and Materials Physics Center MPC, Paseo Manuel de Lardizabal 5, E-20018 San Sebasti\'an, Spain}
\alsoaffiliation{Sainsbury Laboratory, University of Cambridge, 47 Bateman Street, Cambridge CB2 1LR, United Kingdom}

\author{Lorenzo Rovigatti}
\affiliation{Dipartimento di Fisica, Universit\`{a} di Roma La Sapienza, Piazzale Aldo Moro 2, IT-00185 Roma, Italy}
\alsoaffiliation{CNR-ISC Uos Sapienza, Piazzale Aldo Moro 2, IT-00185 Roma, Italy}

\author{Emanuela Zaccarelli}
\affiliation{CNR-ISC Uos Sapienza, Piazzale Aldo Moro 2, IT-00185 Roma, Italy}
\alsoaffiliation{Dipartimento di Fisica, Universit\`{a} di Roma La Sapienza, Piazzale Aldo Moro 2, IT-00185 Roma, Italy}

\author{Francesco Sciortino}
\affiliation{Dipartimento di Fisica, Universit\`{a} di Roma La Sapienza, Piazzale Aldo Moro 2, IT-00185 Roma, Italy}

\author{Angel J. Moreno}
\affiliation{Centro de F\'{i}sica de Materiales (CSIC, UPV/EHU) and Materials Physics Center MPC, Paseo Manuel de Lardizabal 5, E-20018 San Sebasti\'an, Spain}
\alsoaffiliation{Donostia International Physics Center (DIPC), Paseo Manuel de Lardizabal 4, E-20018 San Sebasti\'{a}n, Spain}

\email{angeljose.moreno@ehu.es}
\phone{+34 943 01 8845}

\title[Gel Formation in Reversibly Cross-linking Polymers]{Gel Formation in Reversibly Cross-Linking Polymers}

\abbreviations{SCNP, MD, VMD}

\begin{document}

\begin{abstract}

By means of Langevin dynamics simulations, we investigate the gel formation of randomly functionalized polymers in solution, with the ability to form both intra- and intermolecular reversible bonds. Under highly dilute conditions, these polymers form soft nano-objects (so-called single-chain nanoparticles, SCNPs), resulting from the purely intramolecular cross-linking of the reactive functional groups. Here we show that the competition between intra- and intermolecular bonds at finite concentration is governed by a delicate balance of various entropic contributions and leads to a density dependent effective valence. System-spanning networks are formed at relatively low monomer densities and their stability is mediated by just a small number of intermolecular connections per chain. 
The formation of intermolecular bonds furthermore can induce a non-monotonic dependence of the polymer size on the density for long bond lifetimes. Concomitantly, the polymers in the percolating cluster adopt an intramolecular structure characteristic for self-avoiding chains, which constitutes a strong contrast to the fractal globular behavior of irreversible SCNPs in crowded solutions with purely topological interactions (no intermolecular bonds). Finally, we study the dynamics of the system, which displays signatures expected for reversible gel-forming systems. An interesting behavior emerges in the reorganization dynamics of the percolating cluster. The relaxation is mostly mediated by the diffusion over long distances, through breaking and formation of bonds, of chains that do not leave the percolating cluster. Regarding the few chains that are transiently free, the time they spend until they reattach to the cluster is solely governed by the bond strength.

\end{abstract}

\section{I. Introduction}

Single-chain nanoparticles (SCNPs) are soft nano-objects synthesized from a linear polymer precursor, which is functionalized with reactive groups capable of forming intramolecular bonds. They are typically a few nanometers ($\leq$ 20 nm) in size and possess a large surface-to-volume ratio. The necessary technological ingredients for their synthesis include controlled polymerization, monomer functionalization and cross-linking protocols that ensure their purely intramolecular collapse. Design and function of SCNPs
is a rapidly growing area of research due to their promising applications as catalysts, drug delivery vehicles, biosensors or rheology modifying agents
\cite{Lyon2015,Gonzalez-Burgos2015,Liu2015,Altintas2016,Hanlon2016,PomposoSCNPbook2017,RubioCervilla2017rev,DeLaCuesta2017rev,Kroger2018,Rothfuss2018,Kroger2019,Frisch2020,VerdeSesto2020,Chen2020}. 
Since its beginnings at the start of the 21st century, the synthesis of SCNPs has been dominated by polymer chemistries involving irreversibly cross-linking monomers. In the past years, however, the possibility of exploiting reversible interactions to produce stimuli-responsive SCNPs has gained increased interest \cite{Gonzalez-Burgos2015, Mavila2016, Hanlon2016, Altintas2016, Lyon2015, Sanchez-Sanchez2014, Artar2014, Altintas2012}. Separate classes of reversible interactions with distinct advantages have emerged in the field of single-chain technology: non-covalent and so-called dynamic covalent bonds. Non-covalent bonds are characterized by their relatively low energy (typically a few $k_BT$), which is modulated smoothly by external variables such as temperature, pH and solvent. Prominent examples of non-covalent interactions used in SCNPs are hydrogen bonds \cite{Seo2008, Foster2009}, helical \cite{Terashima2011, Hosono2012} or $\pi-\pi$ stacking \cite{Burattini2009}, host-guest interactions \cite{Appel2012, Wang2016}, ionic attraction \cite{Huh2006} and metal complex formation \cite{Hofmeier2005, Gohy2003}. In contrast to non-covalent bonds, dynamic covalent bonds are very robust and their formation, breaking or exchange can be induced rapidly by very specific external stimuli. These can be pH, photons, redox potentials or a catalyst. The classical example of a dynamic covalent bond is the disulfide bridge, which plays a prominent role in the stabilization of the native folded state of proteins. It served as inspiration for including disulfide bonds \cite{Tuten2012}, but also hydrazone \cite{Murray2011}, enamine \cite{Sanchez-Sanchez2014a}, coumarine \cite{He2011} and anthrazene \cite{Frank2014} bonds in the SCNP chemistry toolbox. 

The advantage of dynamic covalent bonds is that the need for an external stimulus to catalyze their formation and breakage opens up the possibility of kinetically trapping the system. Furthermore, reversibility means that synthesis is never `complete' and individual SCNPs of this kind can form intermolecular bonds in addition to their intramolecular bonds if their concentration is increased above the limit of high dilution. Such intermolecular bonds could potentially lead to aggregation or phase separation but also to the formation of a physical gel. The interplay between intra- and intermolecular bond formation has been exploited recently by Fulton \textit{et al.} in thermoresponsive polymers to produce a system that reversibly crosses between a SCNP solution and an hydrogel \cite{Whitaker2013}. The thermoresponsive nature of the oligoethyleneglycol methyl ether branches causes the polymers to aggregate upon a rise in temperature, while a mildly acidic pH allows the acylhydrazone bonds to undergo component exchange processes. The combination of these two orthogonal triggers leads to the reversible reorganization of intramolecularly folded SCNPs into a robustly cross-linked hydrogel. The response of this material to multiple external stimuli could be exploited in situations where the behavior of the material should depend on the specific makeup of the environment, for example the specific release of drugs in target tissues. 

While the advances in synthesis of such reversible gels made from dynamic covalent SCNPs are promising, investigations of the structure of such materials has been lacking until now. Moreover, the theoretical description of physical gels in general (as opposed to permanently cross-linked chemical gels), has come into focus in the soft matter community in relatively recent years. This is due in part to the difficulty of precisely defining the meaning of `gel' (systems exhibiting both dynamical arrest and network formation are generally considered as gels). A common working definition of a gel is a low density disordered state with solid-like properties such as a yield stress. It combines properties of a liquid (through its disordered structure) and a solid (it does not flow). What distinguishes them from glasses is not only their typical low volume fraction but also their retention of quasi-ergodicity on all but the largest length scales dictated by the infinite percolating network \cite{Zaccarelli2007, Corezzi2003, DelGado2003, Saika-Voivod2004, Rovigatti2011}. A second obstacle for the establishment of a unifying theoretical framework of gel formation is the lack of an ideal model system that incorporates the minimal, necessary ingredients to reproduce the universal features of a gelling system.

In Hill's formalism of liquid condensation in terms of physical clusters, phase separation induced by strong attractive interactions can be avoided by either complementing the attraction by a long-range repulsion \cite{Groenewold2004, Sciortino2005,Ruiz2020} or by modifying the attraction by limiting the valence of the interacting molecules \cite{Zaccarelli2005}. The former can be induced by excessive surface charges on colloids \cite{Segre2001}, while the latter can be achieved, e.g., by decorating colloidal particles with a small number of well defined attractive patches \cite{Manoharan2003, Cho2005} or the engineering of specific DNA sequences designed to form star-shaped architectures with sticky ends \cite{Biffi2013, Rovigatti2014}. An advantage of such limited-valence particles lies in the possibility of theoretically calculating their free energy within the formalism of Wertheim theory \cite{Wertheim1984, Wertheim1984a}, which allows one to determine the phase diagram of the system \cite{Bianchi2006}. Furthermore, the increased experimental control over such patchy particles achieved in the past decade has paved the way for their use as highly tunable building blocks for the design of self-assembled materials \cite{Whitesides2002, Glotzer2004}. 

%\textcolor{red}{shorten the two previous paragraphs?}

Single-chain nanoparticles with reversible bonds may display characteristics of both microphase separating colloids and patchy particles due to the competition between intra- and intermolecular bonds. At very high dilution, intermolecular bonds should be disfavoured by the long intermolecular distances. Upon increasing the volume fraction, some of the intramolecular bonds will be exchanged for connections with other chains for entropic reasons, possibly forming a system-spanning network for the right combination of system parameters. We expect phase separation of the system to be confined to very small densities through the combination of excluded volume interactions and the inherently limited `valence' of the polymers that originates from the locally small number of (monovalent) monomers capable of forming bonds. 
With these ideas in mind, in this article we present Langevin dynamics simulations of solutions of a bead-spring model for
SCNPs with reversible bonds, exploring
concentrations from high dilution to far beyond the overlap density. We characterize the structural and dynamic changes produced by the competition between intra- and intermolecular bonding as the concentration increases.  We find that
the number of intramolecular bonds is not constant but decreases with the concentration, which complicates the possibility of considering the SCNPs as objects with a characteristic valence, and thus of straightforwardly applying Wertheim's formalism. A system-spanning cluster is formed above the overlap concentration. The scaling of the cluster size distribution is consistent with non-mean field critical percolation. The connectivity of the percolating cluster is mediated by a few intermolecular connections per chain. In contrast with the fractal ('crumpled') globular conformations adopted by irreversible SCNPs (no intermolecular bonds) under purely steric crowding
and topological constraints \cite{Moreno2016}, 
intermolecular bonding in the dense solutions just leads to weak perturbations with respect to the limit of high dilution. 
The analysis of the dynamics reveals that, in analogy to the case of colloids with limited valence, the system forms a reversible gel with no apparent signatures of phase separation in the broad range of investigated densities. Relaxation of the dynamic network is mostly mediated by diffusion,  through breaking and formation of bonds, of chains that, without detaching from the percolating cluster, are able to explore distances of several times their size. The article is organized as follows. Section II presents the model and simulation details. Section III shows and discusses structure and dynamics in the scenario
emerging as the concentration increases and the system changes from a dilute solution of SCNPs to a reversible gel. 
Conclusions are given in Section IV.

\section{II. Simulation Details}

The polymers are simulated as chains of beads and springs according to the model of Kremer and Grest \cite{Kremer1990}. 
As such, they represent uncrossable flexible chains with excluded volume interactions in implicit good solvent conditions,
and the bead size $\sigma$ qualitatively corresponds to a Kuhn length \cite{Rubinstein2003} ($\sigma \lesssim 1$~nm). The chains consist of $N=200$ bead (`monomers'), consecutively linked together by irreversible backbone bonds modeled via the FENE potential \cite{Kremer1990}:
\begin{equation}
U^{\rm FENE}(r) = - \epsilon K_{\rm F} R_0^2 \ln \left[ 1 - \left(  \frac{r}{R_0}\right)^2 \right] \, , 
\label{eq:fene}
\end{equation}
with $K_{\rm F} = 15$ and $R_0 = 1.5$. Monomer excluded volume interactions are given by a purely repulsive Lennard-Jones potential (Weeks-Chandler-Andersen potential):
\begin{equation}
U^{\rm LJ}(r) = 4\epsilon \left[ \left(\frac{\sigma}{r}\right)^{12} -\left(\frac{\sigma}{r}\right)^{6} +\frac{1}{4}\right] \, ,
\label{eq:LJ}
\end{equation}
with a cutoff distance $r_{\rm LJ} = 2^{1/6}\sigma$, at which both the potential and the corresponding forces are continuous. 
The sum of both potentials leads to a deep minimum at $r_{\rm min} \lesssim \sigma$ for the total interaction
between two bonded beads, which limits the bond fluctuations and guarantees uncrossability.
The units of energy, length, mass and time are, respectively, $\epsilon, \sigma, m$ and  $\tau = \sqrt{\sigma^2 m/\epsilon}$.
In the rest of the article all numerical values will be given in reduced units $\epsilon = \sigma = m = \tau =1$.

A fraction $f= N_{\rm r}/N$ of the monomers is randomly chosen to be of the reactive type, which can form monofunctional reversible bonds with other reactive monomers.  A bond is formed whenever two unbonded reactive monomers approach each other in space and are separated by less than the capture radius $r_{\rm c} = 1.3$. If there are several candidates within the capture radius a random choice is made.
To avoid trivial bonding, the random distribution of reactive monomers along the chain backbone is made with the constraint that any two consecutive reactive monomers are separated by at least one non-reactive monomer. This bond formation is identical to the cross-linking process in the case of irreversible SCNPs employed in previous studies\cite{Moreno2013, Moreno2016, Formanek2017, Formanek2019, Formanek2019a}. Once a bond is formed, the two participating monomers interact via a Morse potential
\begin{equation}
U^{\rm rev}(r) = K\epsilon\left[2e^{(r_0-r)}-e^{2(r_0-r)}\right] \, , 
\label{eq:morse2}
\end{equation} 
with adjustable parameters $K$ and $r_0$. In the case of the irreversible cross-linking, if two reactive monomers form a bond they interact for the rest of the simulation via the same FENE potential (Eq.~\ref{eq:fene}) as the permanent backbone bonds. 

\begin{figure}[ht]
\centering
  \includegraphics[width=0.65\linewidth]{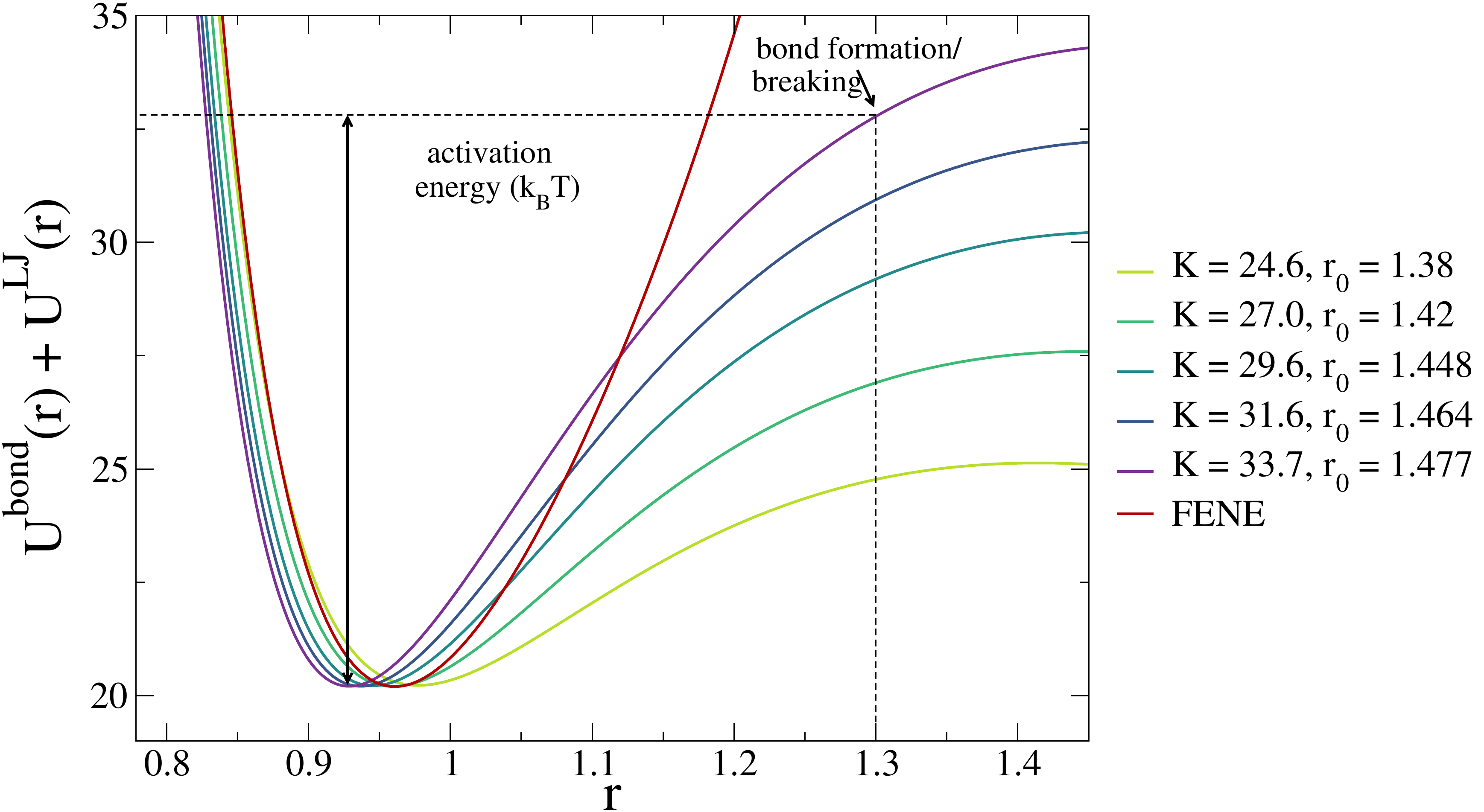}
  \caption{Sum of the bonding ($U^{\rm bond} = U^{\rm FENE}$ for irreversible bonds or $U^{\rm rev}$ for reversible ones) and monomer excluded-volume ($U^{\rm LJ}$) potentials. Data are shown for different bond strengths $K$ and for the irreversible (FENE) case.}
  \label{fgr:potentials}
\end{figure}

Reversible bonds can be broken again if, at any given time step, the participating monomers are separated by 
a distance $r > r_{\rm c}$, upon which their interaction via the Morse potential terminates. The parameter $K$ governs the bond strength through modulating the energy barrier that has to be overcome in order to break the bond. Since $r_{\rm c}> r_{\rm LJ}$ and hence $U^{\rm LJ}(r_{\rm c})=0$, the barrier is given by the energy difference $U^{\rm rev}(r_{\rm c})-U_{\rm min}$, 
where $U_{\rm min}$ is the minimum of $U^{\rm LJ}+U^{\rm rev}$. As such, bond formation is independent of $K$, while bond breakage depends on $K$. Thus, varying $K$ does not only change the average bond lifetime, but also the average probability of any reactive monomer being bonded at equilibrium. The remaining free parameter $r_0$ is chosen such that the minimum $U_{\rm min}$ of the sum of the non-bonded and bonded interactions is,
with respect to the irreversible case,
the same in the energy and as close as possible in the distance (just a slight variation in the range $ 0.93 < r_{\rm min} < 0.98$ is obtained, see Figure~\ref{fgr:potentials}). Contrary to patchy particle models, in which the monofunctionality of the bonds is encoded in the geometry of the interaction \cite{Bianchi2006, Smallenburg2013, Rovigatti2014, Locatelli2017}, we enforce monofunctionality by keeping a list of bonded pairs. Reactive monomers that are mutually bonded cannot form other bonds until their mutual bond is broken (i.e., until their mutual distance becomes $r > r_{\rm c}$).

We perform Langevin dynamics simulations at a fixed temperature  $T=\epsilon/k_{\rm B} = 1$ (with  $k_{\rm B}$ the Boltzmann constant). We use a time-step  $\Delta t = 0.01$ and a friction coefficient  $\gamma = 0.05$. 
The equations of motion are integrated following the scheme of Ref.~\citen{Izaguirre2001}.
Before simulating reversibly cross-linking chains at various densities, we perform exploratory simulations of isolated single chains (mimicking the limit of high dilution), for various values of $f$ (fraction of reactive groups) and $K$ (bond strength). We calculate, for such isolated chains, the radius of gyration $R_{\rm g0}$ and the bond probability $p_{\rm B 0}$, which is defined as the ratio between the average number of formed bonds and the maximum number of bonds (corresponding
to the fully cross-linked SCNP). Our results are summarized in Table \ref{tab:revchainparams}. As expected, the probability of any reactive monomer to be bonded at equilibrium depends more strongly on $K$ than on $f$. 
%Interestingly, a comparison of the radius of gyration of fully cross-linked irreversible SCNPs and the reversible chains of different $K$ shows that SCNPs are still significantly smaller on average than the reversible chains whose bond probability approaches 1. 

Since we are interested in exploring the possibility of forming gels from these reversibly cross-linking chains, the bonding probability needs to be high enough for a system-spanning cluster to form and give the material the ability to propagate stress throughout the whole system. On this basis we choose the bond strength parameters $K=29.6$ and $33.7$ and two fractions of reactive monomers, $f=0.1$ and $0.3$, for the subsequent simulations at different densities. The number of reactive groups in the chain of $N=200$ monomers for such values of $f$ is even, so that in the limit of fully intramolecularly cross-linked SCNPs there are no unbonded reactive groups.

In the solutions of reversible chains we keep the total number of polymers $N_{\rm p}=108$, and thus the total number of monomers $N_{\rm m} =  N_{\rm p} N = 21600$, fixed for all systems.  
We vary the side $L_{\rm box}$ of the cubic simulation box to obtain the desired values of the monomer density
$\rho = N_{\rm m}/L_{\rm box}^3$. The behavior of these systems will be compared with previous simulations of solutions of irreversible SCNPs \cite{Moreno2016,Gonzalez-Burgos2018}. For these simulations $N_{\rm p} =200$ topologically polydisperse, purely intramolecularly cross-linked SCNPs with permanent bonds were first generated. Solutions of such 200 SCNPs were prepared by placing them in the simulation box, preventing concatenations, and slowly compressing the box to different target concentrations, followed by equilibration. Since by construction the intramolecular bonds in the irreversible SCNPs were permanent and no intermolecular bonds were allowed, the intermolecular interactions in these solutions were just given by excluded volume and non-crossability of the chain segments (topological constraints). These interactions are at the origin of the collapse of the irreversible SCNPs to crumpled globular conformations  in crowded solutions \cite{Moreno2016}, instead of the milder transition from self-avoiding to Gaussian conformations adopted by simple unbonded linear chains. For comparison, we also simulated solutions of simple linear chains without reactive groups and with the same $N=200$. Further details of the simulations of the irreversible SCNPs and the linear chains can be found in Refs.~\cite{Moreno2016,Gonzalez-Burgos2018}.
Their gyration radii $R_{\rm g0}$ at $\rho \rightarrow 0$ are included in Table~\ref{tab:revchainparams}. 
\\
\renewcommand{\arraystretch}{1.0}
\renewcommand{\tabcolsep}{13pt}
\begin{table}[ht]
\centering

\begin{tabular}{llllll } 
\toprule
\thead{$f$} & \thead{$K$} & \thead{$r_0$} & \thead{$R_{\rm g0}$} & \thead{$p_{\rm B 0}$} & \thead{$\rho^{\ast}$}\\ \midrule
\multirow{1}{*}{\bf{0.0}} & Linear chains &   & 11.4 & 0    & 0.017 \\ \midrule
\multirow{6}{*}{\bf{0.1}}      & 24.6 & 1.380 & 11.3 & 0.27 & 0.017 \\
                               & 27.0 & 1.420 & 10.9 & 0.47 & 0.019 \\
                               & 29.6 & 1.448 & 10.4 & 0.70 & 0.022 \\
                               & 31.6 & 1.464 &  9.9 & 0.83 & 0.026 \\
                               & 33.7 & 1.477 &  9.3 & 0.91 & 0.031 \\
	    & \multicolumn{2}{l}{Irreversible} & 8.0 & 1    & 0.049 \\ \midrule
\multirow{6}{*}{\bf{0.2}}      & 24.6 & 1.380  & 11.0 & 0.48 & 0.019 \\
			       & 27.0 & 1.420  & 10.4 & 0.68 & 0.022 \\
			       & 29.6 & 1.448  &  9.8 & 0.84 & 0.027 \\
			       & 31.6 & 1.464  &  9.4 & 0.92 & 0.030 \\
			       & 33.7 & 1.477  &  8.8 & 0.96 & 0.037 \\
	    & \multicolumn{2}{l}{Irreversible} &  7.7 & 1    & 0.055 \\ \midrule	
\multirow{6}{*}{\bf{0.3}}      & 24.6 & 1.380  & 10.7 & 0.60 & 0.020 \\
			       & 27.0 & 1.420  & 10.2 & 0.78 & 0.024 \\
			       & 29.6 & 1.448  &  9.7 & 0.90 & 0.027  \\
			       & 31.6 & 1.464  &  9.2 & 0.95 & 0.032 \\
			       & 33.7 & 1.477  &  8.5 & 0.97 & 0.041 \\
	   & \multicolumn{2}{l}{Irreversible}  &  7.5 & 1    & 0.059 \\ \bottomrule
%	   & \multicolumn{2}{l}{Linear chains} &  11.4 & 0 \\ \bottomrule
%\multirow{1}{*}{\bf{0.0}} & Linear chains & & 11.4 & 0 \\ \bottomrule
\end{tabular}
%\end{adjustbox}
\caption{Radius of gyration $R_{\rm g 0}$ and bonding probability $p_{\rm B 0}$, as a function of the bond strength $K$ and the fraction of reactive groups $f$, at highly dilute conditions ($\rho \rightarrow 0$). The overlap density $\rho^{\ast}$ (in number of monomers per volume) is included. The values of all the former quantities for irreversibly cross-linked SCNPs and for simple unbonded linear chains are included for comparison.}
\label{tab:revchainparams}
\end{table}

For the remainder of the article, we will use the term `solutions of SCNPs' only to refer to the solutions of irreversible SCNPs without intermolecular bonds. The reversible systems will be denoted by their values of $K$ and $f$. The density of the system will be given in reduced units $\rho/\rho^{\star}$, where $\rho^{\star}$ is the overlap concentration. We define this quantity as $\rho^{\star}= N(2R_{\rm g0})^{-3}$, with $R_{\rm g0}$ the radius of gyration of the isolated polymer ($\rho \rightarrow 0$, see Table \ref{tab:revchainparams}). We expect intermolecular cross-links to begin forming significantly around the overlap concentration, when monomers of different chains start to enter the same space. 
%Irreversible SCNPs at equilibrium undergo a crossover in their scaling behavior around the overlap concentration. Their topological interactions prevent the concatenation of two or more SCNPs, which leads to their collapse to `crumpled' or fractal globules \cite{Moreno2016,Gonzalez-Burgos2018}, instead of the well-known milder transition of linear polymers to Gaussian chains \cite{Rubinstein2003}. 
We explore densities in the range $0 \leq \rho/\rho^{\star} \leq 4.4$. This corresponds to monomer densities up to $\rho \approx 0.14$, which lies below the entanglement density for linear chains of the same polymerization degree, $\rho_{\rm e} \gtrsim 0.42$. The latter is calculated as \cite{Rubinstein2003} 
$\rho_{\rm e} = (N_{\rm e}/N)^{3\nu_{\rm F}-1}$, where $\nu_{\rm F} = 0.588$ is the Flory exponent and $N_{\rm e}$ 
is the entanglement length of linear chains in the melt state. For the simulated bead-spring model $\rho \approx 1$ corresponds to the
melt state \cite{Kremer1990} and a value $N_{\rm e} \gtrsim 65$ is found \cite{Sukumaran2005}.
For each combination ($K, f$) and each density 8 independent simulation runs were carried out, each consisting of $1 \times 10^7$ equilibration steps and $4 \times 10^7$ production steps.

%For comparison, we include the behavior of irreversibly cross-linked SCNPs (no intermolecular bonds) under crowding conditions, whose collapse behavior is much more pronounced and resembles that of ring polymers  \cite{Moreno2016}. For the remainder of the article, whenever we refer to SCNPs for comparison, we mean topologically polydisperse solutions of single-chain nanoparticles, which were obtained by an irreversible cross-linking procedure at effective infinite dilution, such that no intermolecular bonds are present in the system. Namely, isolated linear precursors were coupled to the same Langevin thermostat with intermolecular interactions
%switched off, so that cross-linking was intramolecular by construction. Dilute solutions of the irreversibly cross-linked SCNPs generated in that way were
%constructed by placing them in the simulation box, at mutual distances long enough to avoid intermolecular contacts and concatenations.
%The box was slowly compressed and equilibrated at the target densities. Further details can be found in Refs.~\cite{Moreno2016,Gonzalez-Burgos2018}

\section{III. Results and Discussion}

\subsection{Competition between intra- and intermolecular bonds}\label{CompetitionIntraInter}
Ideally, one would wish to derive a thermodynamic description of the system, for example according to Wertheim theory \cite{Wertheim1984, Wertheim1984a}, by using inputs from computer simulations, which would allow us to evaluate the complete $K - \rho$ phase diagram of these reversibly cross-linking polymers, and to find the regions in which gel formation is possible. Wertheim thermodynamic perturbation theory (TPT) was originally developed for associating liquids, but has also been successfully employed to elucidate the phase behavior of gel forming systems of limited valence, qualitatively and sometimes even quantitatively reproducing numerical results \cite{Bianchi2006, Roldan-Vargas2013, Marshall2012}. 

A few fundamental assumptions of TPT have to be satisfied in order to be able to describe the system according to its predictions: (i) bonds are strictly monofunctional, (ii) two molecules cannot share more than one bond and (iii) molecules cannot form bonds with themselves. While the assumption (ii) might be problematic at very high densities, (iii) is inherently violated in a flexible polymeric molecule with many functional groups along its backbone. However, if the number of intramolecular bonds stays approximately constant at different densities, we can neglect intramolecular bonds and view the polymers as having an `effective' valence of $M = [1-p_{\rm B }]N f$. If this is the case, the polymers might behave similar to patchy particles, but with the distinction that the `patches' are not located at specific points on their surface, but randomly distributed, and fluctuate due to the inherent softness of the polymer and the rearrangement of the intramolecular bonds along the chain backbone. In this view, the intramolecular bonds solely affect the reference free energy.

%\begin{figure}[hbp]
%\centering
 % \includegraphics[width=0.7\linewidth]{{Gels/Intramol_change}.eps}
%  \caption{Relative change in average number of intramolecular bonds per molecule (with respect to infinite dilution) for different values of energy constant $K$ and reactive monomer fraction $f$ as a function of density.}
%  \label{fgr:intramolbonds}
%\end{figure}

\begin{figure}[ht]
%\centering
%\begin{minipage}[t]{0.48\textwidth}
%\begin{subfigure}[b]{0.45\textwidth}
\centering
\includegraphics[width=0.52\textwidth]{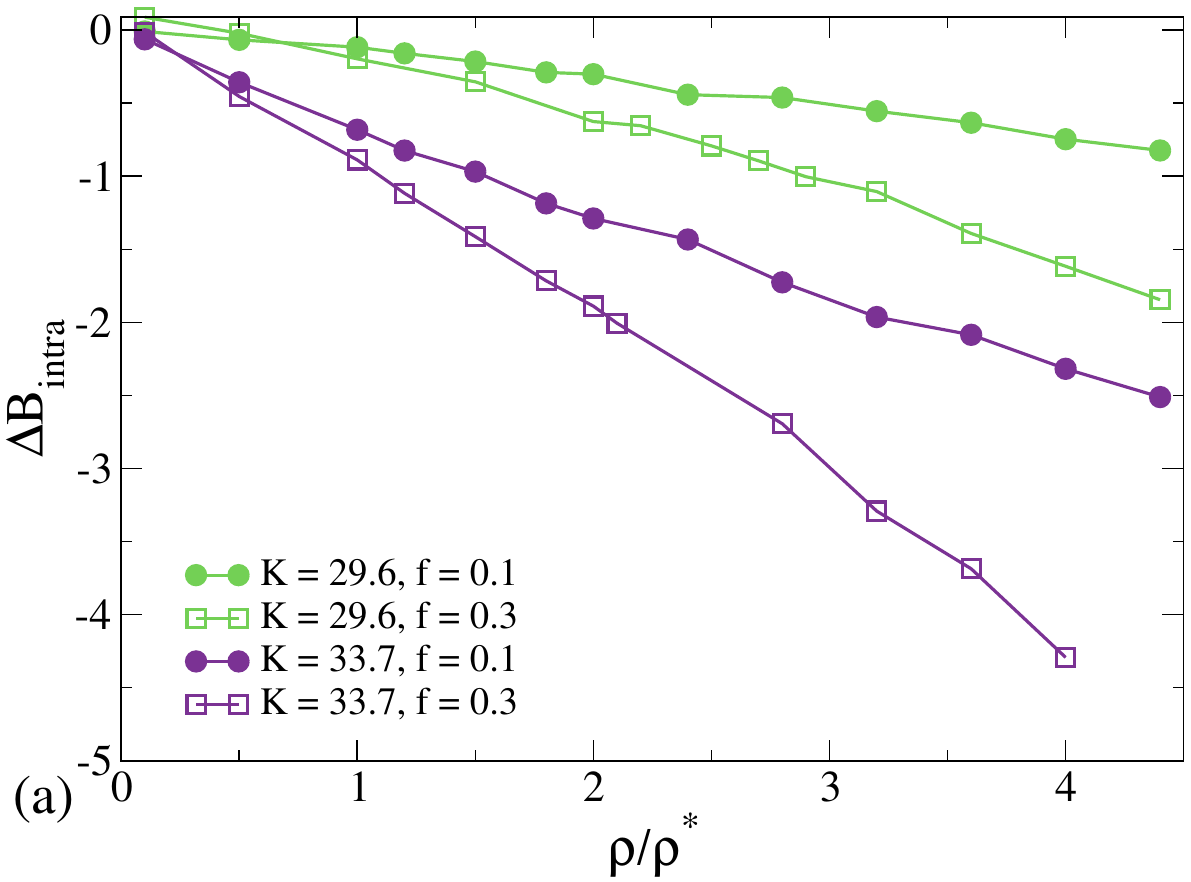}
\includegraphics[width=0.54\textwidth]{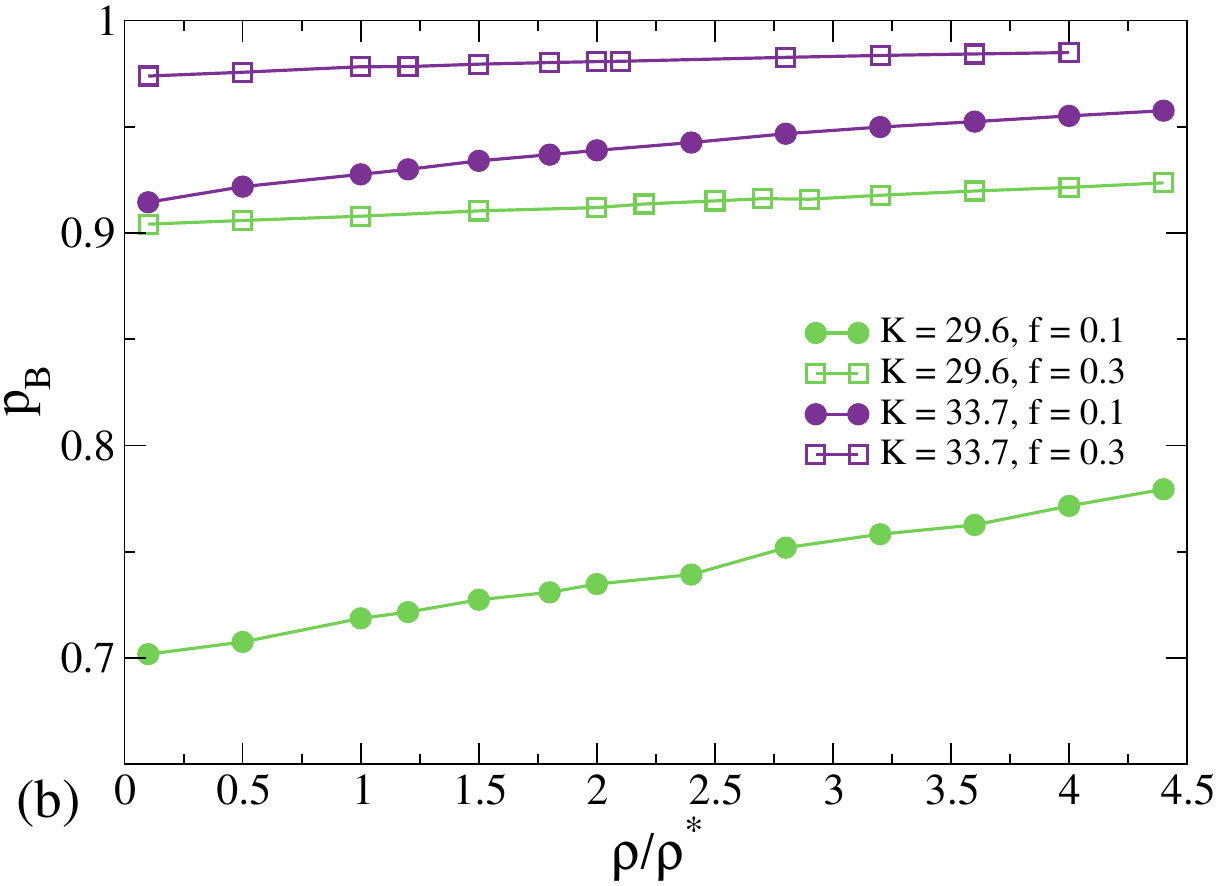}
\caption{(a) Change in average number of intramolecular bonds per molecule with respect to infinite dilution, for different values of the energy constant $K$ and reactive monomer fraction $f$, as a function of the density.
(b) Total (sum of intra- and intermolecular) bonding probability $p_{\rm B}$ for different values of $(K,f)$, as a function of the density.}
\label{fgr:totalbonds}
\end{figure}

%\label{fgr:intramolbonds}
%\end{figure}
%
%\end{minipage}%
%\hfill
%\end{subfigure}
%\begin{subfigure}[b]{0.45\textwidth}
%\end{figure}
%\begin{figure}[h!]
%\begin{minipage}[t]{0.48\textwidth}%
%
%\begin{figure}
%\centering
%\includegraphics[width=0.7\textwidth]{{Gels/freact}.eps}
%\caption{Total (intra- and intermolecular) bond probability $p_B$for different values of energy constant $K$ and reactive monomer fraction $f$ as a function of density.} 
%\label{fgr:totalbonds}
%\end{minipage}
 %\end{subfigure}

We test whether the assumption of the intramolecular bonds being unaffected by interactions with other polymers holds over a certain range of densities.  Figure~\ref{fgr:totalbonds}a displays the loss of intramolecular bonds per chain as a function of the density. For all ($K, f$) parameter combinations, increasing the density of the system leads to a competition between intra- and intermolecular bonds, instead of a simple addition of intermolecular bonds on the periphery of an intramolecularly cross-linked polymer. The effect is strongest for $K=33.7$ and $f=0.3$, the system with the highest bonding probability $p_{\rm B 0}$ at high dilution. This can be understood in terms of the total (intra- and intermolecular) bonding probability $p_{\rm B}$, which is displayed in Figure~\ref{fgr:totalbonds}b as a function of the density. We notice that at $K=33.7$ and $f=0.3$ the bonding probability is already saturated at low density and stays almost constant upon crowding the system. On the other hand, the parameter combination with the smallest bonding probability at high dilution, $K=29.6$ and $f=0.1$, exhibits the strongest increase in $p_{\rm B}$ as a function of density. Here, the loss in entropy stemming from crowding of the surrounding molecules can be largely compensated by the enthalpic gain of forming new additional intermolecular bonds.
Since Figure~\ref{fgr:totalbonds}a shows that the number of intramolecular bonds is not constant but strongly decays as the density grows, we conclude that a straightforward application of Wertheim's theory is not possible in the present case.

\begin{figure}[ht]
\centering
\includegraphics[width=0.45\linewidth]{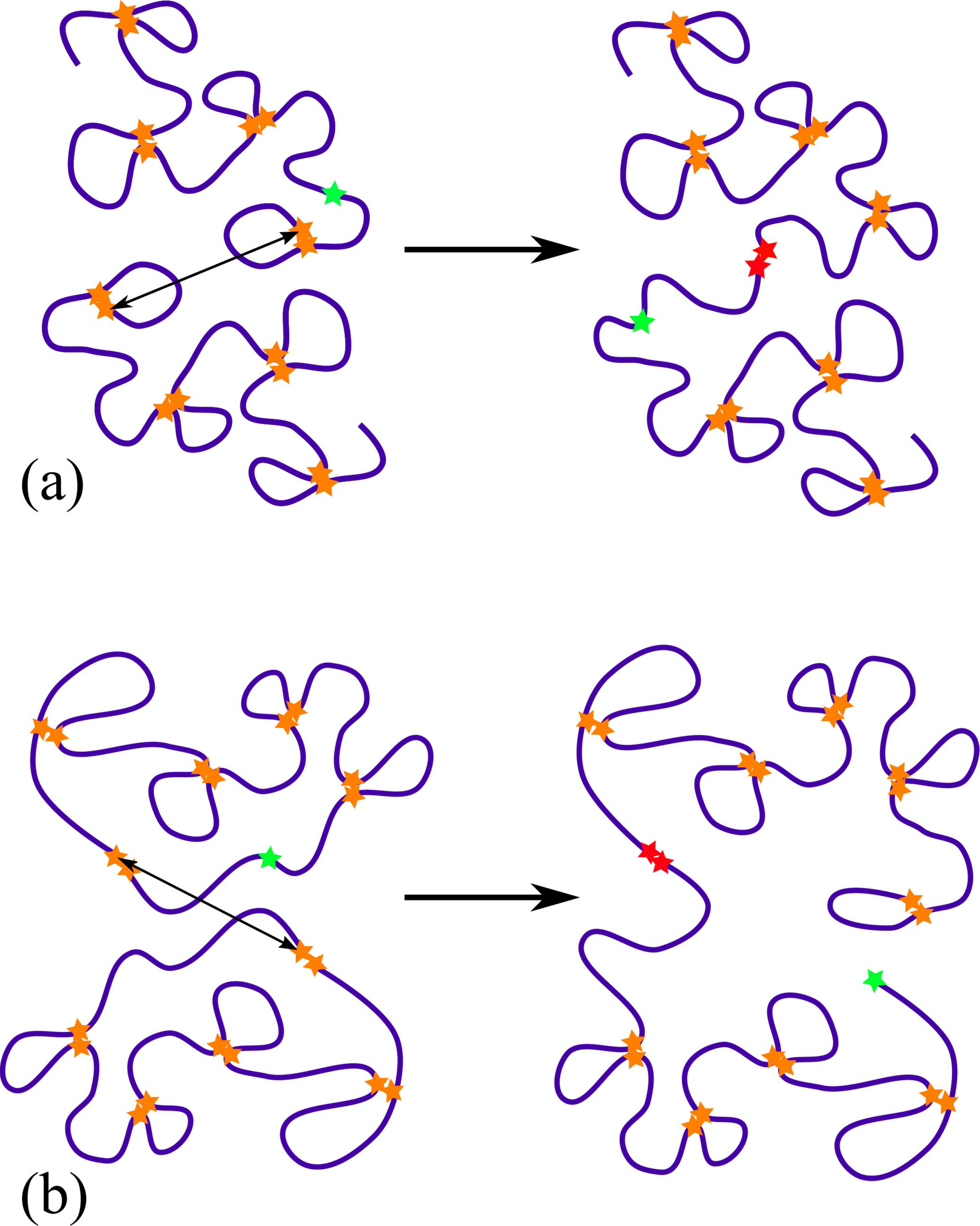}  
  \caption{Schematic examples of two possible recombinations between intramolecular (orange stars) and intermolecular (red stars) bonds. Both the number of free reactive monomers (green stars) and the total number of bonds remains unchanged in both events. (a): two short intramolecular loops are exchanged for an intermolecular bond; (b): two long intramolecular loops  break through the formation of an intermolecular bond. The total number of bonds remains unchanged in both (a) and (b) through the additional formation of a small intramolecular loop.}
  \label{fgr:bond-exchange}
\end{figure}

The competition between intra- and intermolecular bonds when the system is highly bonded has to be understood in terms of a delicate interplay between various entropic and energetic contributions to the free energy of the system. One might expect that the formation of intramolecular bonds is favourable over the formation of connections with other molecules, as the latter reduces the translational entropy of both molecules without a compensating energetic gain (the bonds are energetically equivalent). On the other hand, depending on the contour length separating the monomers whose intramolecular bond is exchanged for an intermolecular bond, it might be that the breakage of the former could potentially increase the conformational entropy of the molecule that loses the intramolecular link. Figure \ref{fgr:bond-exchange} presents an example of two possible bond recombination events. In scenario (a), two intramolecular bonds between monomers separated by short contour distances are exchanged for an intermolecular bond (and a different intramolecular bond to keep the number of free reactive groups constant). The conformational entropy is thus only mildly affected by the breakage of the intramolecular bonds, while the translational entropy is decreased significantly, as the two molecules now have to diffuse together. In scenario (b), each separate molecule contains a long-range loop formed by the connection of its two backbone ends. The opening of this loop via the exchange for an intramolecular bond increases the conformational entropy of both molecules, as one of their ends (the one not participating in the newly formed intermolecular bond) becomes floppier and acquires increased freedom to explore different conformations. Apart from the entropic contributions of increasing or decreasing conformational and translational degrees of freedom, one has to consider the purely combinatorial increase in entropy due to the possibility of forming intermolecular bonds \cite{Sciortino2019,Sciortino2020}. This delicate interplay between the various entropic contributions should be at the origin of the loss of intramolecular bonds in favour of energetically equivalent intermolecular bonds when the concentration increases, and  renders a treatment within Wertheim theory impossible without significant modifications of the theory. 

\subsection{Intermolecular Bonding and Percolation}
A necessary, albeit not sufficient, prerequisite for gelation is the emergence of a fully connecting network, spanning the whole system in all three directions. In chemical gels, where bonds are irreversible, the onset of this percolation coincides with the system acquiring a finite shear modulus and an infinite zero shear viscosity  (the gel stops flowing). In physical gels, where bonds are transient, clusters can break and reform over time, which strongly affects the mechanical and dynamical properties of the system. The transient appearance of a system spanning cluster does therefore not guarantee the propagation of external stresses throughout the whole system for all time scales, as would be expected in a gel. 
The latter is only possible if an infinite cluster persists in time.

\begin{figure}[ht]
\centering
  \includegraphics[width=1\linewidth]{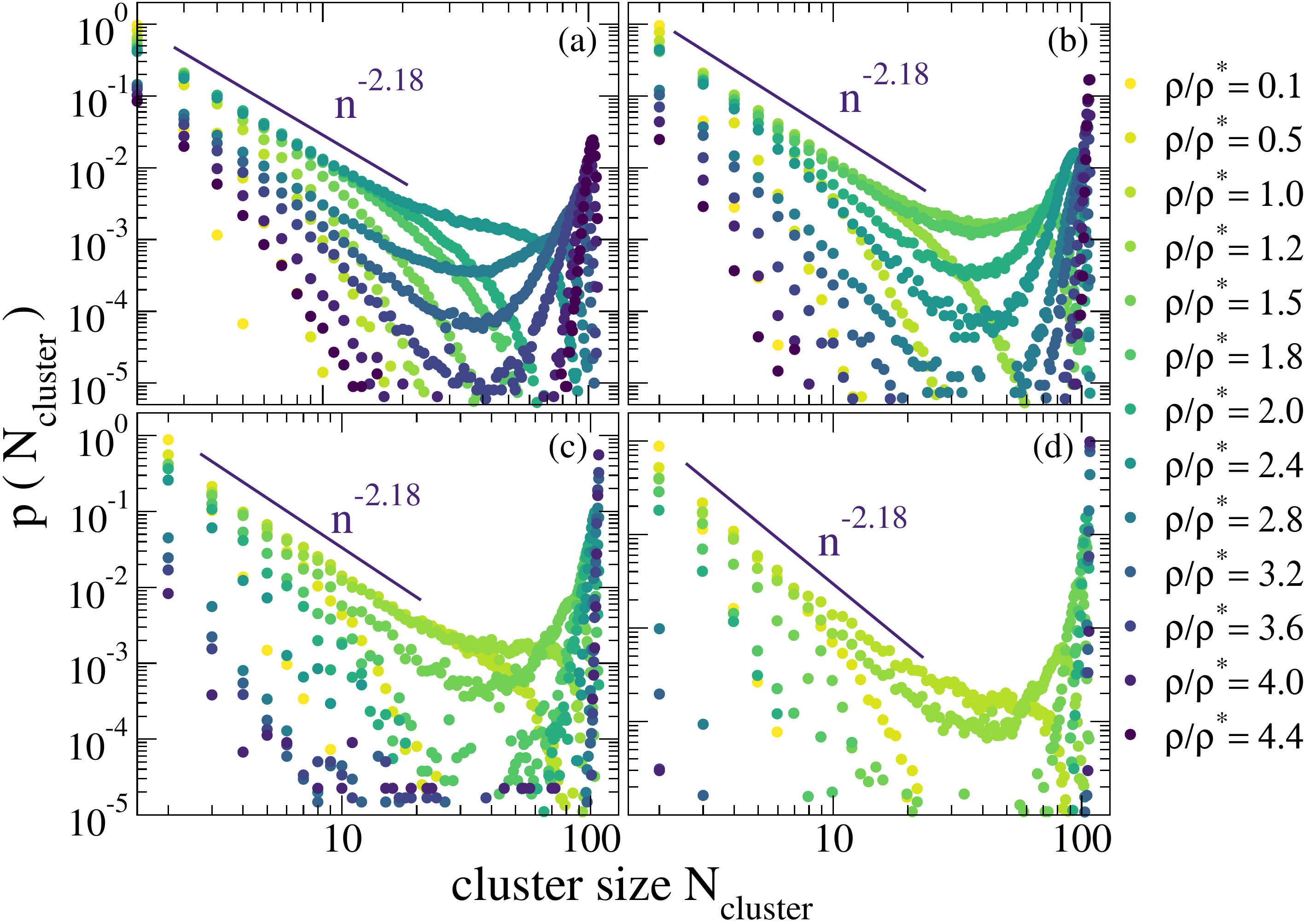}
  \caption{Cluster size distribution for $K=29.6$ (a, b) and $K=33.7$ (c, d) with $f=0.1$ (a, c) and $f=0.3$ (b, d) for various densities (see legend). Solid lines are power-laws $p(N_{\rm cluster}) \sim N_{\rm cluster}^{-2.18}$.}
  \label{fgr:percolation}
\end{figure}

We explore the formation of a spanning cluster in our system. 
Figure \ref{fgr:percolation} displays the distribution of cluster sizes for the whole range of densities and all the combinations of the parameters $K$ and $f$. 
Here, two chains  belong to the same cluster if they are mutually connected by at least one intermolecular bond. The maximum cluster size $N_{\rm cluster}^{\rm max}$ is 108, 
i.e., the total number of polymers in the simulation box, which has been kept fixed for all the concentrations. 
Irreversible gelation processes are well described by the Flory-Stockmayer (FS) mean-field theory of percolation if two conditions are met: the bonds are independent from each other and loops are not present in the system \cite{Flory1941, Stockmayer1943, Stockmayer1944}. Under these assumptions, if the functionality of the monomers is known, the percolation threshold can be calculated in terms of a critical bond probability $p_{\rm B}^{\rm c}$, which depends on the temperature or attraction strength and on the volume fraction \cite{Flory1941, Stauffer1992}. Close to this critical point, the cluster size distribution follows a power law $p(N_{\rm cluster}) \sim N_{\rm cluster}^{-\tau}$ with exponent $\tau = 5/2$. In real systems (with cluster loops), scaling properties are retained, but the value of the critical exponent is different. In three dimensions, numerical calculations on different lattices yield an exponent of about $\tau \approx 2.18$ \cite{Stauffer1975, Kirkpatrick1976, Stauffer1979}. As can be seen in Figure \ref{fgr:percolation},  close to the formation of a shoulder in the cluster size distribution, this is well described by a power-law with an exponent $\tau \approx 2.18$. This (non mean-field) result is found consistently for all the combinations of $K$ and $f$.

\begin{figure}[ht]
\centering
%\begin{minipage}[t]{0.48\textwidth}
%\begin{subfigure}[b]{0.45\textwidth}
%\centering
\includegraphics[width=0.55\textwidth]{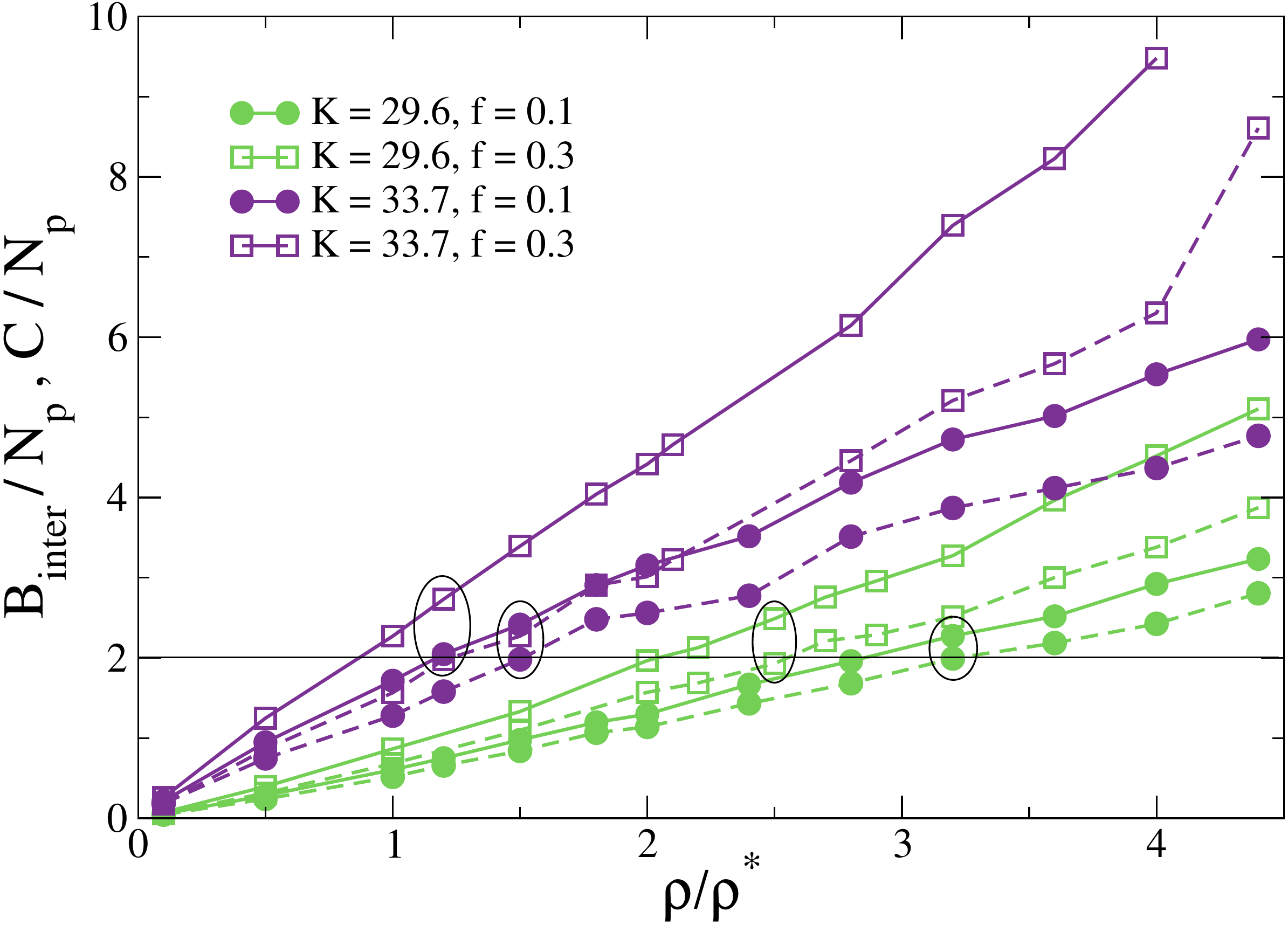}
\caption{Average number of intermolecular bonds ($B_{\rm inter}$, solid lines) and connectivity ($C$, dashed lines) per molecule for different values of the energy constant $K$ and reactive monomer fraction $f$, as a function of the density.
The values at which a pronounced peak at large cluster sizes first appears in $P(N_{\rm cluster})$ 
(see Figure~\ref{fgr:percolation}) are marked by ellipses. }
\label{fgr:intermolbonds}
%\end{minipage}%
\end{figure}

%\hfill
%\end{subfigure}
%\begin{subfigure}[b]{0.45\textwidth}
%\end{figure}
%\begin{figure}[h!]
%\begin{minipage}[t]{0.48\textwidth}%

\begin{figure}[ht]
\centering
\includegraphics[width=0.55\textwidth]{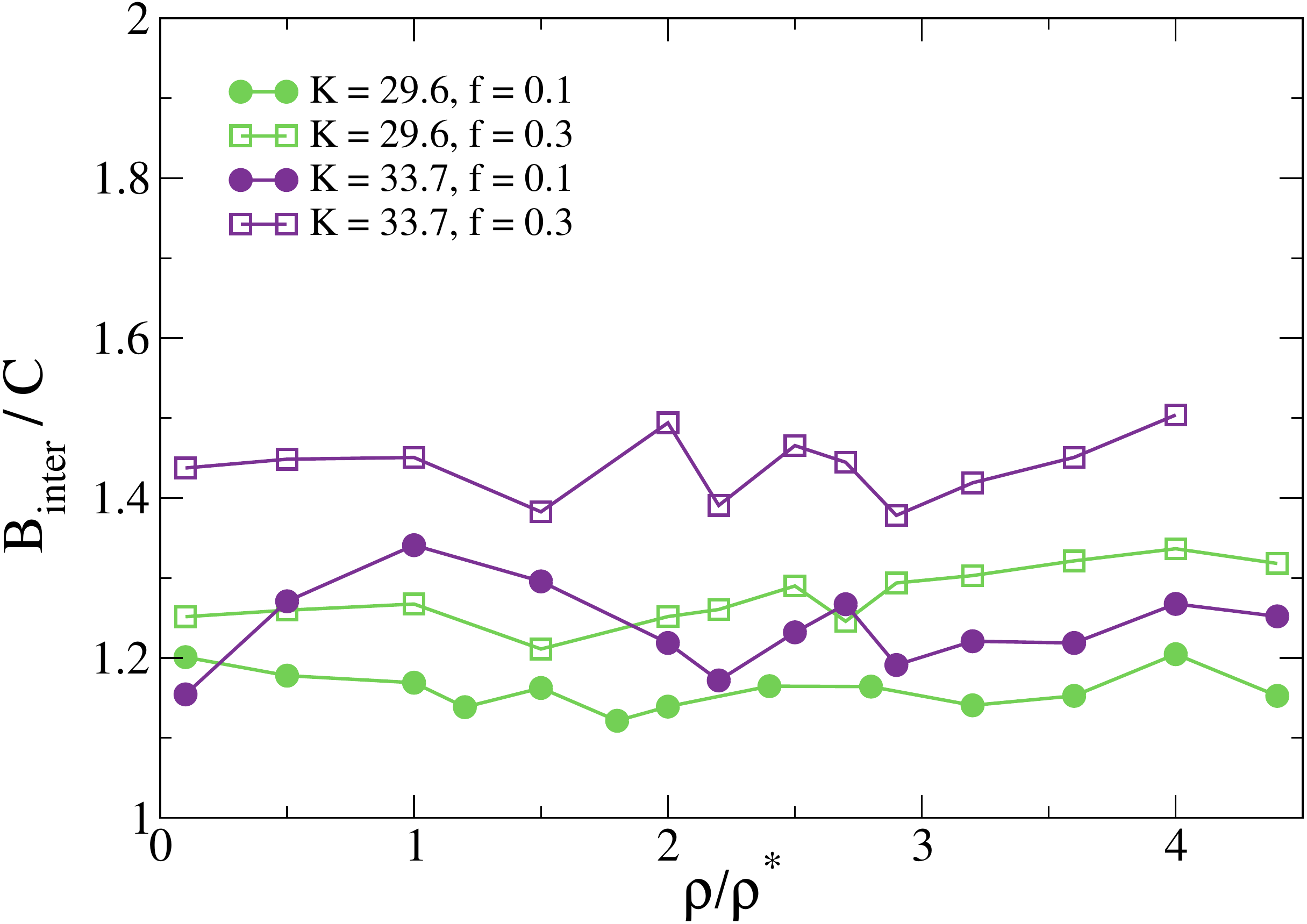}
\caption{Average number of intermolecular bonds $B_{\rm inter}$ per connection with a different molecule, for different values of the energy constant $K$ and reactive monomer fraction $f$, as a function of the density. } 
\label{fgr:bondpercontact}
%\end{minipage}
 %\end{subfigure}
\end{figure}

Since the `effective valence' of these reversibly cross-linking polymers is dependent on the density, reactive monomer fraction and bond strength, we are interested in how many intermolecular bonds a chain forms on average as the system starts to percolate. One has to keep in mind, however, that multiple bonds are possible between two specific polymers, and every additional bond shared between two given chains does not add to the overall connectivity of the network. In Figure \ref{fgr:intermolbonds} we thus display both the average number of intermolecular bonds per chain as well as the average connectivity $C$ (i.e. to how many different chains is a given chain connected through inter-chain bonds, irrespective of the number of bonds mediating the connection between each pair of chains). The values at which a pronounced peak at large cluster sizes first appears in $P(N_{\rm cluster})$ (see Figure \ref{fgr:percolation}) are marked by ellipses in Figure~\ref{fgr:intermolbonds}. These values provide an estimation of the percolation line and, as can be seen, this occurs at densities above and even far above the overlap concentration. 
The general trend is that the percolation transition approaches the overlap concentration from above by increasing the
fraction of reactive groups and the bond strength.

Figure~\ref{fgr:intermolbonds} shows that the average connectivity is $C \approx 2$ 
around the percolation transition. 
%which represents the minimal inter-chain connectivity necessary for such a network to occur. 
%Looking at Figure \ref{fgr:intermolbonds} one might conclude from the increasing difference between the number of intermolecular bonds $B_{\rm inter}$ and the connectivity $C$ at high densities, 
%One might expect that for high densities at some point bonds will start to be formed preferentially between chains that are already linked together. This intuitively makes sense, especially for high fractions of reactive monomers, where they are statistically close in space, facilitating the formation of further bonds once one connection between two polymers is established. 
We calculate the ratio $B_{\rm inter}/C$, i.e., how many intermolecular bonds are formed on average 
between two mutually connected chains (Figure \ref{fgr:bondpercontact}). Interestingly, increasing the concentration does not lead to 
a higher number of bonds per pair of connected chains. The number of interchain connections increases but the average number of bonds mediating
a connection, $B_{\rm inter}/C$, stays approximately constant across the whole range of densities for all the ($K,f$) parameter combinations. Higher fractions of reactive monomers $f$ lead  to higher values of $B_{\rm inter}/C$, as expected. In all cases an average of about 1.2-1.4 intermolecular bonds mediating an interchain connection is sufficient to mantain the spanning network.

\begin{figure}[ht]
\centering
  \includegraphics[width=1\linewidth]{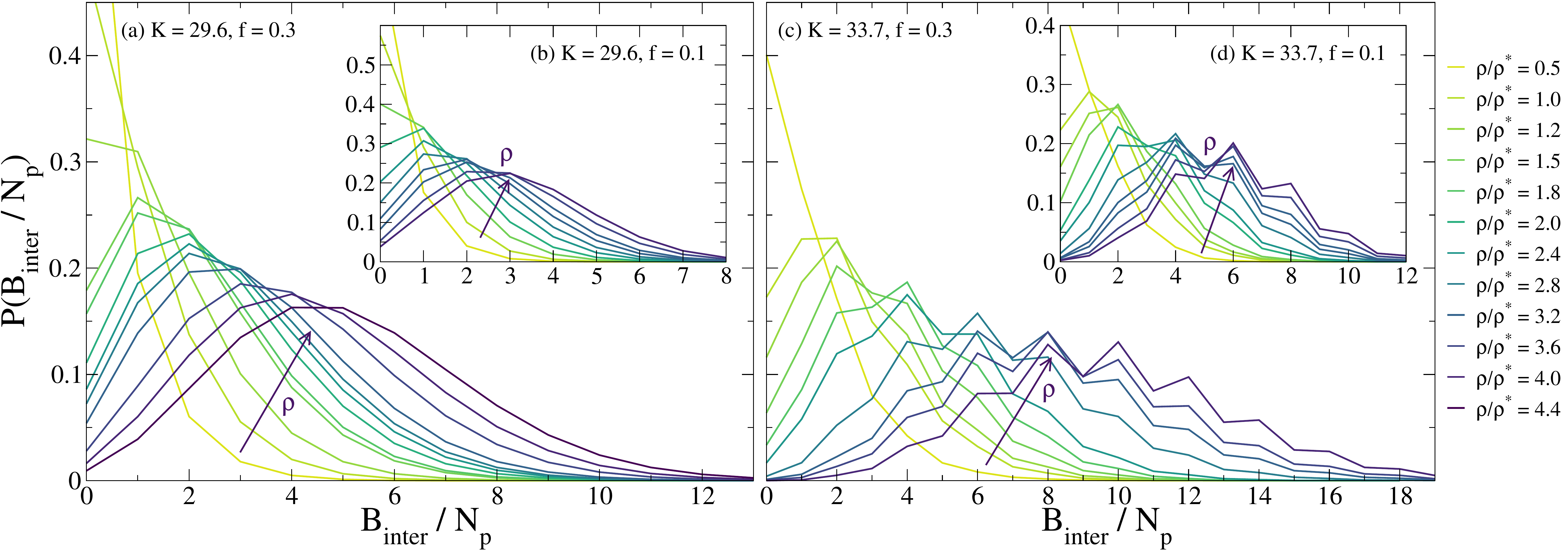}
  \caption{Distribution of intermolecular bonds per chain for $K=29.6$ (a, b) and $K=33.7$ (c, d) for various densities. The main panels depict the case $f=0.3$ (a, c), while the insets represent the case $f=0.1$ (b, d). The arrows indicate the direction of increasing density (see also legend).}
  \label{fgr:bondhist}
\end{figure}

To gain a deeper understanding of the formation of intermolecular bonds, we next investigate the distribution of intermolecular bonds at specific densities. Figure \ref{fgr:bondhist} displays the distributions for $K=29.6$ and $K=33.7$, the main panels presenting the case $f=0.3$ and the insets the case $f=0.1$. The distributions reveal a difference between the energy landscapes of the systems for both bond strengths. Whereas for $K=29.6$ the distributions are smooth with a clear maximum, the data for $K=33.7$ exhibit a characteristic zig-zag pattern, with even values of $B_{\rm inter}$ being more favourable than their closest odd values. This feature can be understood as follows.  Since, as mentioned in Section II, the total number of reactive groups in the chain is even, forming an odd number of intermolecular bonds means that at least one reactive monomer of the chain remains unbonded. On the other hand, the number of unreacted monomers
is very low for $K=33.7$ since indeed $p_{\rm B}$ is very close to the full-bonding limit $p_{\rm B}=1$ (Figure~\ref{fgr:totalbonds}).
In this limit of saturation the penalty in energy and combinatorial entropy is high enough to disfavour odd numbers 
of intermolecular bonds per chain over even ones. 

\subsection{Structural Properties}

\begin{figure}[ht]
\centering
  \includegraphics[width=0.55\linewidth]{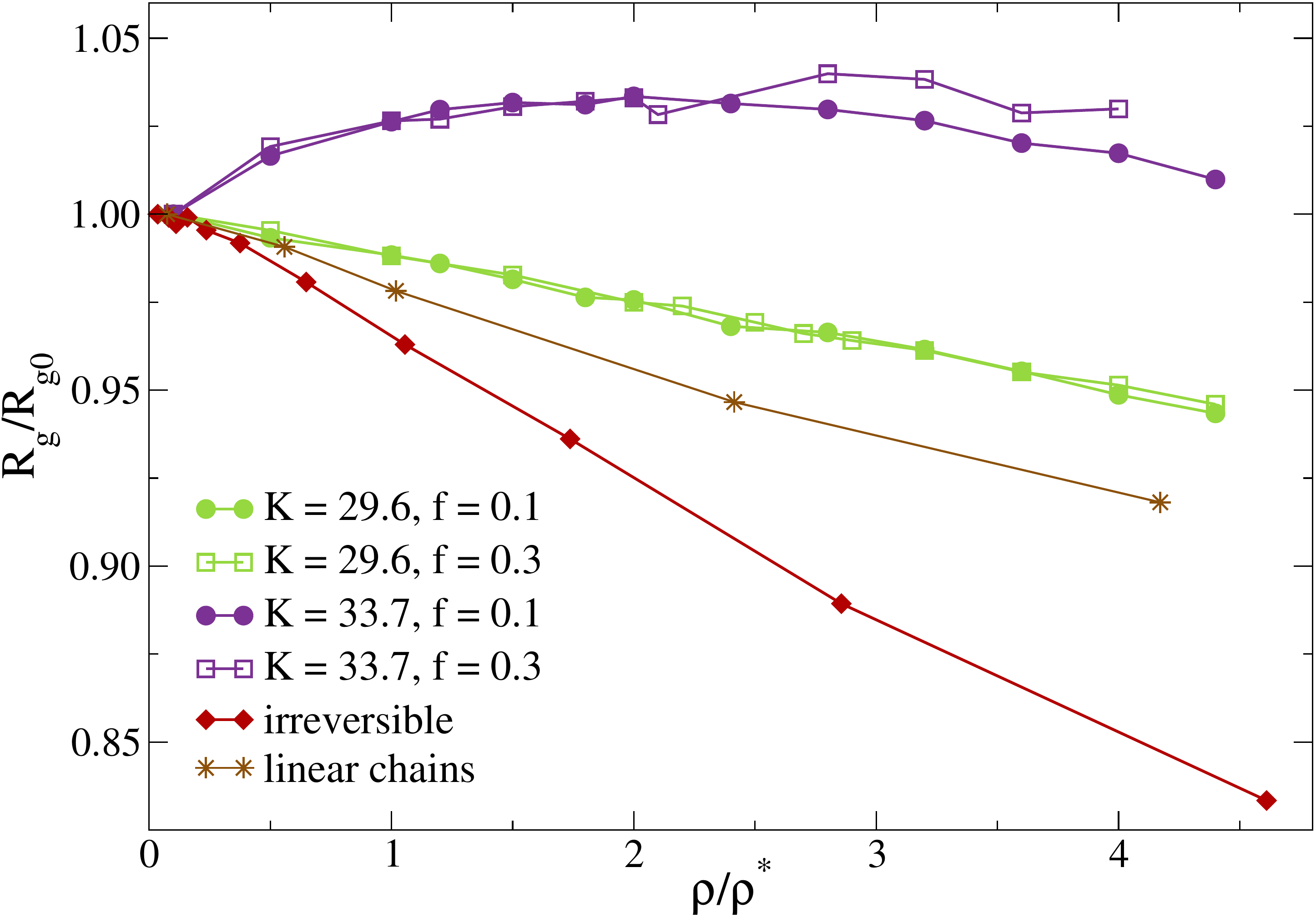}
  \caption{Normalized radius of gyration for different values of the energy constant $K$ and reactive monomer fraction $f$, as well as for irreversible SCNPs (no intermolecular bonds) and simple linear chains without reactive groups, as a function of the density.}
  \label{fgr:Rggels}
\end{figure}

In the previous section, we have shown that the intramolecular bonds are not unaffected by the presence of other molecules, but rather that intra- and intermolecular bonds compete with each other, the outcome of which depends on a delicate interplay of various entropic contributions. We expect this exchange of intra- for intermolecular bonds to be accompanied by structural changes in the polymers. The partial unfolding induced by the opening of intramolecular loops might to some degree counteract the collapse found above the overlap concentration in the case of the irreversible SCNPs.
A first measure of structural changes upon increasing the density is the size of the polymers, given by the radius of gyration depicted in Figure \ref{fgr:Rggels}. We find that the competition between steric repulsion and partial unfolding leads to qualitatively different density dependences of the molecular size. For $K=29.6$, shrinking due to macromolecular crowding dominates, whereas for the strong association energy $K=33.7$, the polymers swell slightly with respect to their conformations at high dilution. At high densities, however, a re-entrance of $R_{\rm g}$ for $K=33.7$ can be observed. For comparison we include 
the corresponding data for linear chains without reactive groups and for the irreversible SCNPs (no intermolecular bonds). In both cases shrinking is stronger 
than in the reversible systems. The effect is especially pronounced in the irreversible SCNPs, resembling the observation of ring polymers collapsing to crumpled globular structures \cite{Halverson2011,Moreno2016,Gonzalez-Burgos2018}.

\begin{figure}[ht]
\centering
  \includegraphics[width=1\linewidth]{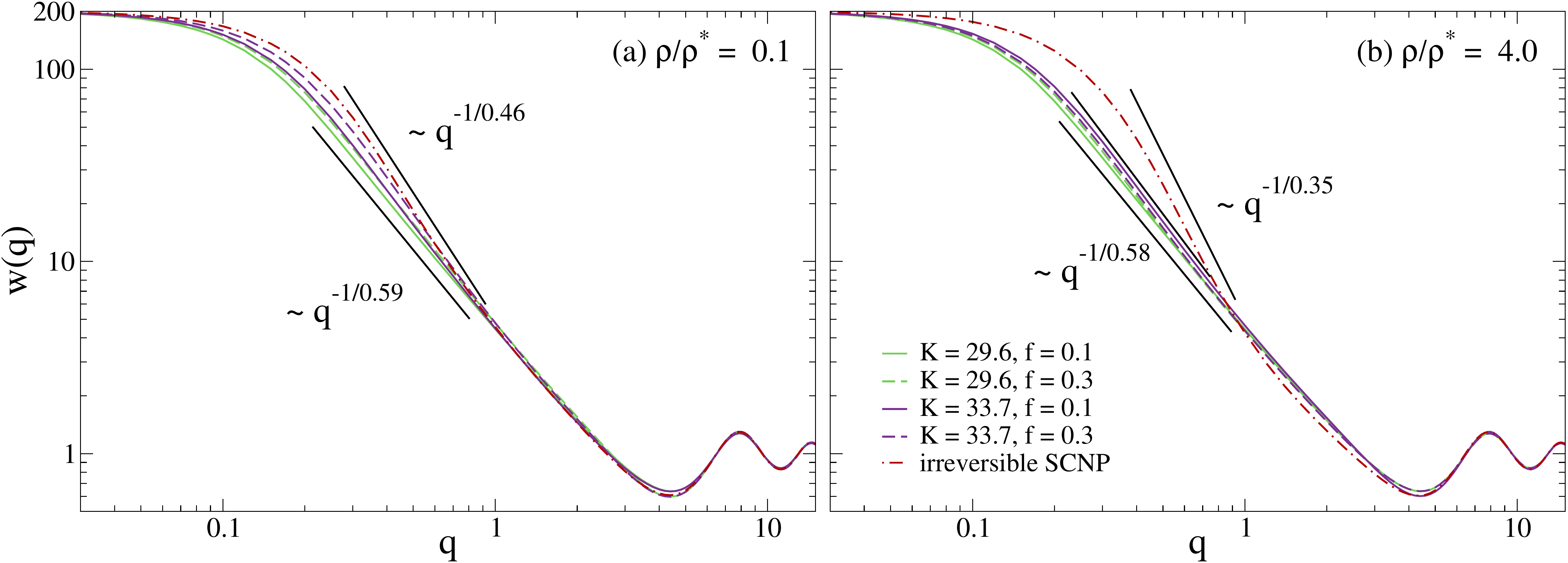}
  \caption{Intramolecular form factor $w(q)$ for all ($K, f$)-parameter combinations at densities  (a) $\rho/\rho^{\star} = 0.1$ and (b) $\rho/\rho^{\star} = 4.0$, below and above the percolation threshold for all parameters, respectively. Dot-dashed lines are data for irreversibly cross-linked SCNPs. Solid black lines are power laws representing the approximate scaling of $w(q)$ in the fractal regime. }
  \label{fgr:formfactor}
\end{figure}

A more detailed description of the intramolecular structure of the polymers is given by their isotropic form factors,  
\begin{equation}
w(q) = \left\langle \frac{1}{N} \sum_{j,k} \frac{\sin(qr_{jk})}{qr_{jk}} \right\rangle  \, , 
\label{eq:formfactor-iso2}
\end{equation}
where $r_{jk}$ is the distance between monomers $j$ and $k$ and the sum only includes pairs of monomers belonging to the same molecule. Being fractal objects, the form factors of polymers typically follow a scaling law $w(q) \sim q^{-1/\nu}$ for wave vectors corresponding to length scales bigger than the bond length $b$, but smaller than the radius of gyration, i.e. $1/R_g \lesssim q \lesssim 1/b$. Figure \ref{fgr:formfactor} displays the form factors for all $(K,f)$ parameter combinations at a low density ($\rho/\rho^{\star}=0.1$, panel (a)) and a high density  beyond the percolation threshold 
($\rho/\rho^{\star}=4.0$, panel (b)). Results for the irreversible SCNPs are included for comparison.
Far below the overlap concentration, the scaling exponents adopted by the chains with a higher bonding probability $p_{\rm B0}$ at $\rho \rightarrow 0$ (higher $K$ and $f$) are systematically lower than for those with lower $p_{\rm B0}$ (lower $K$ and $f$). At such concentrations the intramolecular bonds are clearly dominant and a higher number of bonds (higher $p_{\rm B}$)
leads to a stronger compaction (lower $\nu$) of the polymers, though the effect is moderate (changing from $\nu = 0.59$ for the weakest bonds to
$\nu = 0.46$ for the permanent ones). This trend is consistent with experimental scaling exponents for reversible 
and irreversible SCNPs at high dilution \cite{Pomposo2017}.

%concentration \textcolor{red}{This corresponds to the local compaction induced by the intermolecular bonds. EXPLAIN BETTER}. Interestingly, even chains with $K=33.7$ and $f=0.3$, whose bond probability $p_{\rm B} = 0.97$ is very close to 1, still exhibit pronounced differences with irreversible SCNPs (red dot-dashed lines). Indeed, a recent review of experimental data of reversible SCNPs in high dilution showed that their structure is well described by a self-avoiding walk with a Flory-like scaling of $\nu \approx 0.6$, regardless of the specific polymer chemistry or the kind of reversible interaction (e.g. hydrogen bonds, disulfide bridges, complex formation) \cite{Pomposo2017}. This scaling is also found in our system for the lower bond strength $K=29.6$. \textcolor{red}{The difference from the self-avoiding character when the bond life-time approaches infinity (irreversible case) is a kinetic effect that stems from the occasional arrest of a rare conformation through the formation of a long-range loop. EXPLAIN BETTER.} 

Far beyond the percolation threshold, the form factors of the reversible polymers of different bond strength and reactive fraction are nearly indistinguishable. Their scaling exponent $\nu \approx 0.58$ is very close to the Flory exponent for self-avoiding conformations, and clearly
distinct from the exponent $\nu \sim 1/3$ characterizig the crumpling globular conformations of the irreversible SCNPs
in dense solutions. Such a big difference suggests that, when concentrating the solution, the long-range intramolecular loops of the intramolecularly bonded SCNPs are more likely to be exchanged for intermolecular bonds than the short-range ones (as discussed earlier, this exchange can be explained by a gain in conformational entropy via the opening of the long-range loops, see Figure \ref{fgr:bond-exchange}). Moreover, the formation of new long intramolecular loops will involve concatenations with neighbouring similar loops. 
Both mechanisms reduce the presence of the non-concatenated long loops that are at the origin of the crumpled globular 
conformations of the irreversible SCNPs in dense solutions \cite{Moreno2016,Gonzalez-Burgos2018}.   
%We also note that the cross-links in irreversible SCNPs render them less penetrable (concatenation of permanent loops is forbidden), which leads to their collapse to crumpled globules \cite{Moreno2016}, with much lower exponents ($\nu \sim 1/3$) than in the reversible polymers.     

\begin{figure}[ht]
\centering
\includegraphics[width=0.55\linewidth]{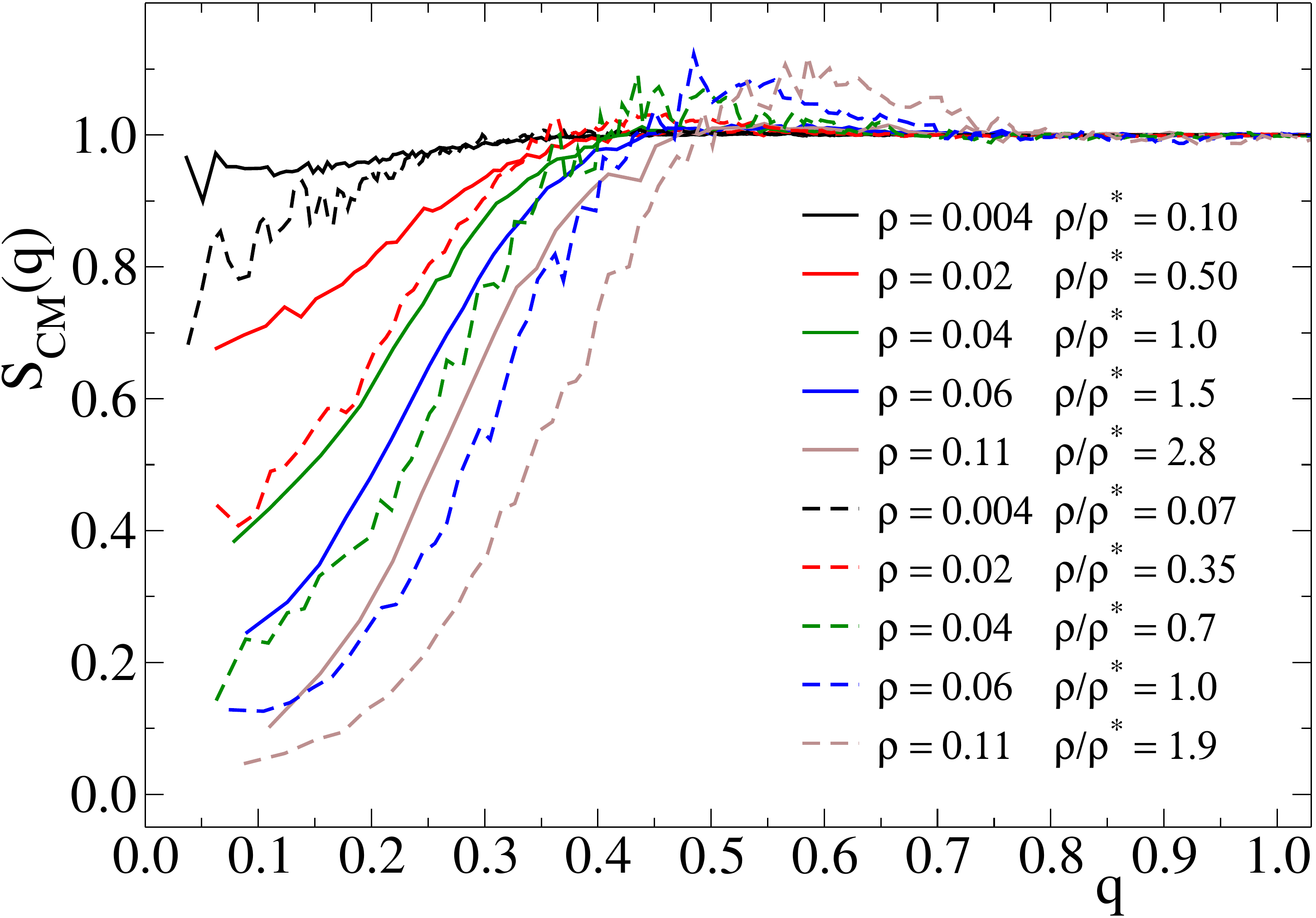}  
  \caption{Structure factors of the macromolecular centers-of-mass for reversible chains (solid lines) with $K=33.7$ and $f=0.3$. 
Dashed lines represent data for the irreversibly cross-linked SCNP counterparts. Same colors correspond to same absolute values ($\rho$) of the 
monomer density. The legend also indicates the normalized density ($\rho/\rho^{\star}$).}
  \label{fgr:structurefactor}
\end{figure}

The large-scale structure of the system can be probed by calculating correlations between the macromolecular centers-of-mass through the structure factor 
\begin{equation}
S_{\rm CM}(\mathbf{q}) = \left\langle \frac{1}{N_{\rm p}} \sum_{j,k} \exp\left[ i \mathbf{q}\cdot ( \mathbf{r_{\rm CM,j}} - \mathbf{r_{\rm CM,k}} )\right] \right\rangle \, , 
\label{eq:structurefactor}
\end{equation}
where $\mathbf{r_{\rm CM,j}}$ denotes the position of the center-of-mass of the macromolecule $j$.
Figure \ref{fgr:structurefactor} shows $S_{\rm CM}(q)$ for reversible chains with $K=33.7$ and $f=0.3$ at various monomer densities. The polymers with other
values of ($K,f$) exhibit qualitatively the same behavior (not shown). For comparison we include the corresponding data for irreversible SCNPs at the same monomer densities. 
%In the intermediate to high-$q$ range, both the reversible intermolecularly bonded polymers and the irreversible purely intramolecularly cross-linked SCNPs exhibit qualitatively the same behavior. 
In the limit of $q \rightarrow 0$ the structure factor of the macromolecular centers-of-mass is proportional to the compressibility $\chi$ of the material
\begin{equation}
\lim_{q\to 0} S_{\rm CM}(q) = \chi \rho k_B T \, . 
\end{equation}
At the same monomer density the low-$q$ limit of $S_{\rm CM}(q)$ is significantly and consistently lower for the solution formed by irreversible SCNPs than for the network made up of reversibly bonded chains. The permanent cross-links in the irreversible SCNPs prevent 
concatenation of loops and lead to less interpenetration of different SCNPs, rendering the system less compressible 
(lower $S(q=0)$ and $\chi$) than the reversible network at the same density. Furthermore, in the investigated range of concentrations
$0.1 < \rho/\rho^{\ast} < 4.4$, the structure factor shows no sign of phase separation or growing inhomogeneities 
(which would be manifested by a diverging or strongly growing $S(q)$ as $q \rightarrow 0$).

\begin{figure}[ht]
\centering
 \includegraphics[width=0.55\linewidth]{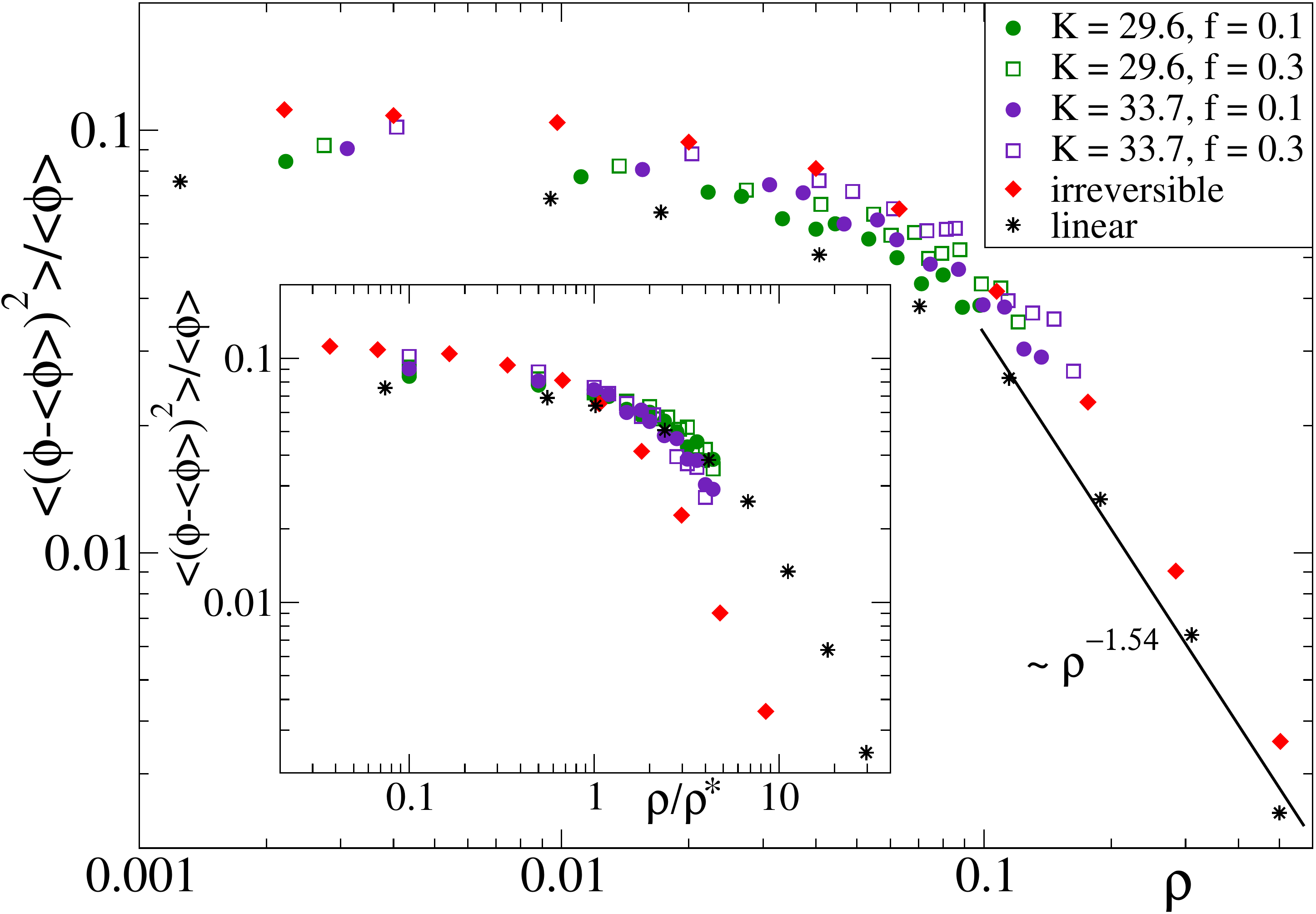} 
  \caption{Average monomer density fluctuation, normalized by $\langle \phi \rangle = \rho$ (see text) for a bin size of $L_{\rm s} = 5$, for different values of the energy constant $K$ and reactive monomer fraction $f$. Results for the irreversible SCNPs and for linear chains are included. Data in the main panel and in the inset are represented vs. the absolute and the normalized ($\rho/\rho^{\ast}$) monomer densities, respectively.}
  \label{fgr:densfluc}
\end{figure}

To assure that no phase separation is indeed intervening with gelation, we investigate density fluctuations in the system. To this end, the box is divided into sub-boxes of side $L_{\rm s}$ and the `local' monomer density $\phi$ is calculated within each of them. Density fluctuations are then defined by the parameter
%\begin{equation}
$\delta\phi^2 = \langle \left(\phi - \langle \phi \rangle \right)^2 \rangle $, 
%\end{equation}
%     
where the average is both taken over all sub-boxes and over various realizations (therefore the mean value of $\phi$
is identical to the macroscopic monomer density, $\langle \phi \rangle = \rho$). 
In the absence of correlations, fluctuations just behave as \cite{BarratHansen2003} $\delta\phi^2 \sim \rho$.
Figure~\ref{fgr:densfluc} shows the normalized quantity  $\delta\phi^2 /\rho$ as a function of the concentration,
so that the effect of the correlations is reflected by deviations from a plateau.
The data in the main panel are shown as a function of the absolute monomer density $\rho$.
In order to higlight the different trends found below and above the overlap concentration,
the inset shows the data as a function of $\rho/\rho^{\star}$.
The choice of the sub-box size $L_{\rm s}$ influences the values of $\delta\phi^2 /\rho$. If $L_{\rm s}$ is too small, fluctuations are inherently limited, while if it is too big the phase occupying less volume might not be properly sampled. We varied $L_{\rm s}$ in a reasonable range $ 3 \leq L_{\rm s} \leq 20 $ and the fluctation size showed the same dependence on the concentration irrespective of the specific $L_{\rm s}$, the numerical values just changing by a scaling factor. Data in Figure~\ref{fgr:densfluc}
were obtained by using $L_{\rm s}=5$.

At high dilution, $\rho/\rho^{\ast} \ll 1$, $\delta\phi^2 /\rho$ for the linear chains and the irreversible SCNPs shows a plateau.  
As aforementioned this result reflects vanishing correlations and is fully expected in such systems, due to both high dilution and 
the absence of bonding and aggregation. Above the overlap concentration
($\rho/\rho^{\star} >1$), in semidilute conditions, the low-$q$ limit of the structure factor of linear chains is expected to follow the relation \cite{Rubinstein2003} $S(0)/\rho \sim \delta \phi^2/\rho \sim \xi^2$, where $\xi$ 
is the correlation length. The portion of the linear chain within this length scale keeps its high-dilution scaling, i.e., $\xi \sim g^{\nu_{\rm F}}$, with $g$ the number of monomers of the chain in the correlation volume. Combining the former relations we obtain: 
\begin{equation}
\rho \sim g/\xi^3 \sim \xi^\frac{1 -3\nu_{\rm F}}{\nu_{\rm F}}, 
\end{equation}
and the concentration dependence of the fluctuations follows the power-law:
\begin{equation}
\delta\phi^2/\rho \sim \xi^2 \sim \rho^\frac{-2\nu_{\rm F}}{3\nu_{\rm F}-1}
\end{equation}
Since $\nu_{\rm F}= 0.588$, the fluctations in the solutions of linear chains are expected to decrease 
with the concentration as $\delta\phi^2/\rho \sim \rho^{-1.54}$. This is confirmed in Figure~\ref{fgr:densfluc}.
Similar results are found for the irreversible SCNPs.

Remarkably, the  solutions of reversibly cross-linking polymers show the same qualitative trends 
as the solutions of non-aggregating linear chains or irreversible SCNPs, 
in particular the plateau at low concentrations. Quantitative differences are found in the relative amplitudes of the fluctuations, which can be easily rationalized. For the same values of $L_{\rm s}$ and $\rho$ the amplitude of $\delta\phi^2/\rho$ is systematically higher as the molecular size decreases (see values of $R_{\rm g0}$ in Table~\ref{tab:revchainparams}).
Since all systems have the same molecular weight, smaller macromolecular sizes correspond to more compact objects, which at the same macroscopic density $\rho$ lead to higher values of $\delta\phi^2$ when they are distributed over the simulation box.
In summary, data in Figure~\ref{fgr:densfluc} are consistent with the results for the structure factors of Figure~\ref{fgr:structurefactor} and no signs of phase separation are found (otherwise consistently growing values of $\delta\phi^2/\rho$ with decreasing $\rho$ would be found instead of the plateau). Still, phase separation cannot be fully discarded ---one should have in mind that fluctuations are limited by the simulation box. 
Unfortunately the evaluation of the phase diagram for small densities is quite challenging and 
we cannot give a fully conclusive answer to this issue.

\subsection{Dynamic Properties}

\begin{figure}[ht]
\centering
  \includegraphics[width=0.55\linewidth]{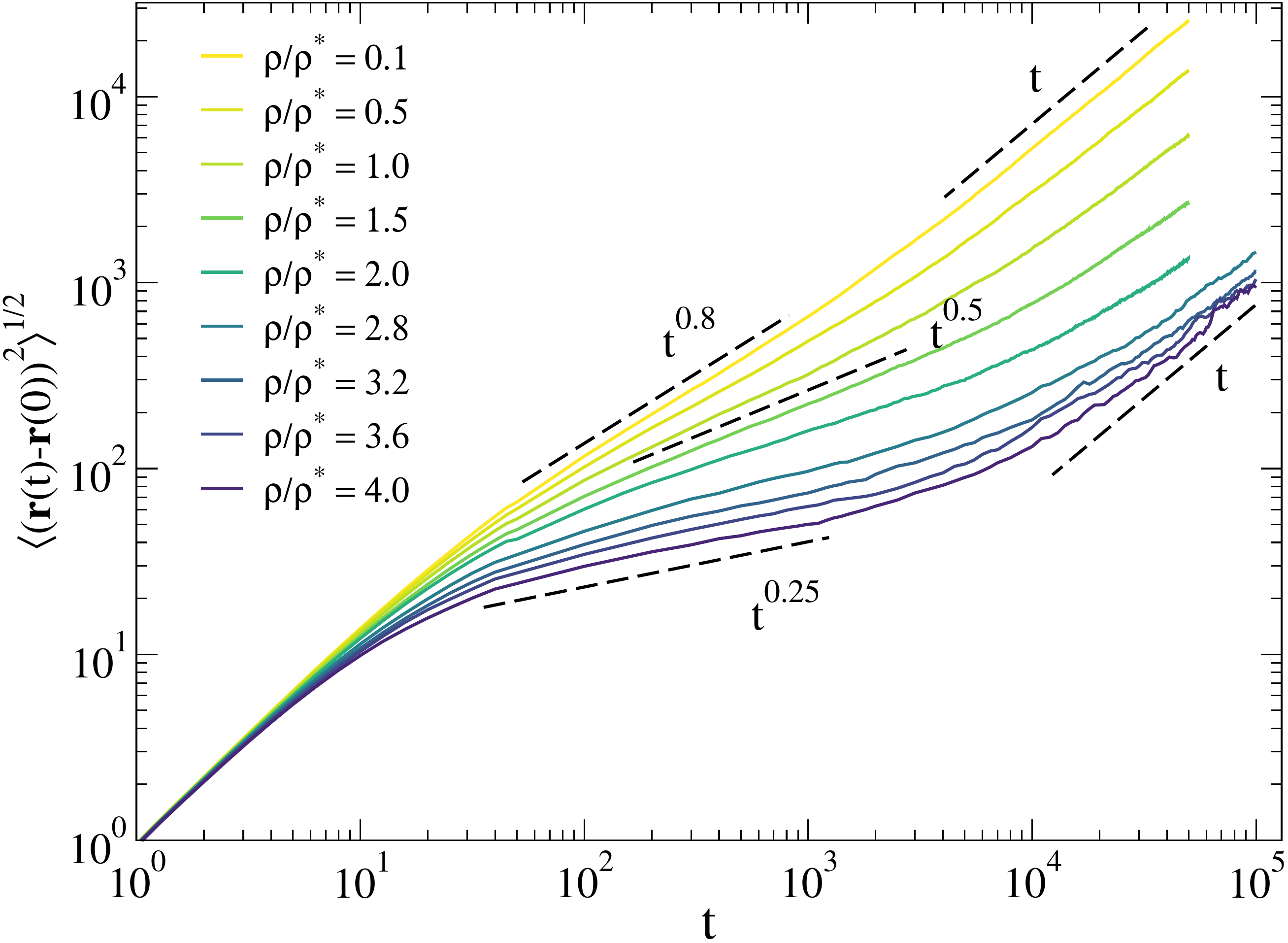}
  \caption{Monomer mean-squared displacement for $K=33.7$ and $f=0.3$ at various densities.
  The dashed lines indicate approximate power-laws.}
  \label{fgr:msd}
\end{figure}

After having discussed in detail the structural properties of the networks formed through the reversible intermolecular bonds, we now shift our focus to the dynamics of the system. First, we calculate the mean-squared displacement of the individual monomers, ${\rm MSD}(t) = \langle \left( {\bf r}_i(t) - {\bf r}_i(0)\right)^2 \rangle $.
Figure~\ref{fgr:msd} displays the MSD for the polymers with $K=33.7$ and $f=0.3$ at various densities. The other investigated reversible polymers display qualitatively the same trends (not shown). At short time-scales, monomers diffuse freely without a density dependence. Upon increasing the density, a `soft' plateau appears in the MSD, marking a slowing down of the dynamics \cite{noterouse}. The soft plateau can be described by a subdiffusive power law, MSD~$\sim t^x$, with an exponent that decays with increasing
the density ($x \approx 0.25$ at the highest simulated concentration). 
%At the highest density considered, this plateau extends up to 2 time decades. 
At high concentrations when localization of the polymers is stronger, the square-root of the value of the MSD in the plateau regime,
which gives a measure of the localization length, is $\Delta=\sqrt{{\rm MSD}} \approx 5-8$, i.e, of the order of the radius of gyration of the polymer. This is quite large, given that on average every third monomer along the chain (since $f=0.3$) is reactive and can potentially be engaged in a bond that strongly limits fluctuations.

\begin{figure}[ht]
\centering
  \includegraphics[width=0.55\linewidth]{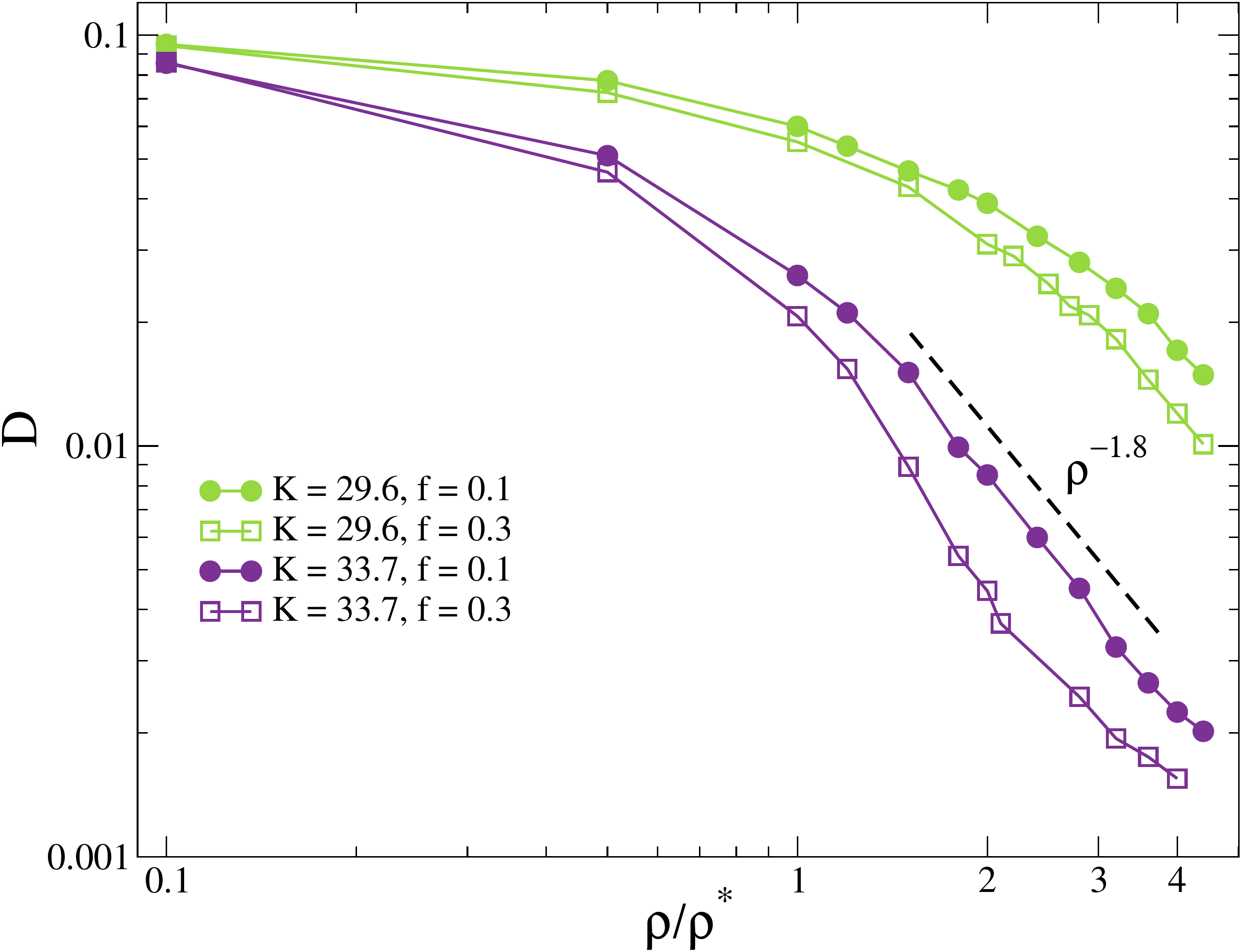}
  \caption{Density dependence of the diffusivity, calculated from the monomer mean-squared displacement in the diffusive regime, for different combinations of $K$ and $f$.}
  \label{fgr:diffusion}
\end{figure}

In all the simulated systems monomers reach the diffusive regime at long time scales, characterized by ${\rm MSD} \sim 6Dt$, with diffusivity $D$. Still, the large localization length and the progressive slowing down reflected in Figure~\ref{fgr:msd} are clear signatures of approaching a gel transition. The values of $D$, obtained as MSD$/6t$ at the end of the simulation time window, are presented in Figure~\ref{fgr:diffusion}
for different parameters ($K,f$). Not surprisingly, the diffusivity shows a sharp decay when the concentration is increased beyond the overlap density and at every time most of the polymers are bonded to the percolating cluster. The decay can be approximately described by a power-law $D \sim \rho^{-1.8}$. Interestingly, in the systems with $K=33.7$ there is an apparent second crossover to a milder dependence on the concentration if this is further increased. A tentative explanation for this feature might be as follows. We anticipate that in this regime almost all the polymers remain linked to the percolating cluster during the whole time scale of the simulation (see below). 
If a chain is linked to the percolating cluster, it will need to break some intermolecular bond to perform a broad fluctuation and partially relax its conformation until it forms a new bond at another point of the network. Full relaxation and diffusion of the chain is achieved through a succession of these events. They are favoured, and partially compensate the decrease of the diffusivity produced by crowding, 
at concentrations for which the distances to unbonded reactive groups in {\it other} chains become comparable or smaller than 
distances between close reactive groups in the same chain. This hypothesis can be supported by an estimation of such distances. 
In that range of high  concentrations, for $K=33.7, f=0.3$ there is typically one unreacted monomer per chain. Therefore the typical distance to the closest unreacted monomer in another chain can be estimated as $d_{\rm inter} \sim L_{\rm box} N_{\rm p}^{-1/3} \sim 2R_{\rm g0}(\rho/\rho^{\ast})^{-1/3}$. By using the $R_{\rm g0}$
values in Table~\ref{tab:revchainparams}, and the range $\rho/\rho^{\ast} \sim 2-4$ where we observe the crossover to the mild regime in the diffusivity, the former 
distance is $d_{\rm inter} \sim 13-15$. On the other hand there are in average $N_{\rm p}/3$ monomers between consecutive reactive monomers in a same chain (since $f=0.3$), which corresponds to a distance in the real space of $d_{\rm intra} \sim (C_{\infty} N_{\rm p}/3)^{\nu}$. For the used bead-spring model the characteristic ratio \cite{Rubinstein2003} is $C_{\infty} \approx 1.7$, 
and at high concentrations $\nu = 0.58$ (Figure~\ref{fgr:formfactor}b).
Therefore $d_{\rm intra} \sim 16$, which is comparable or larger than $d_{\rm inter}$.

\begin{figure}[ht]
\centering
  \includegraphics[width=1\linewidth]{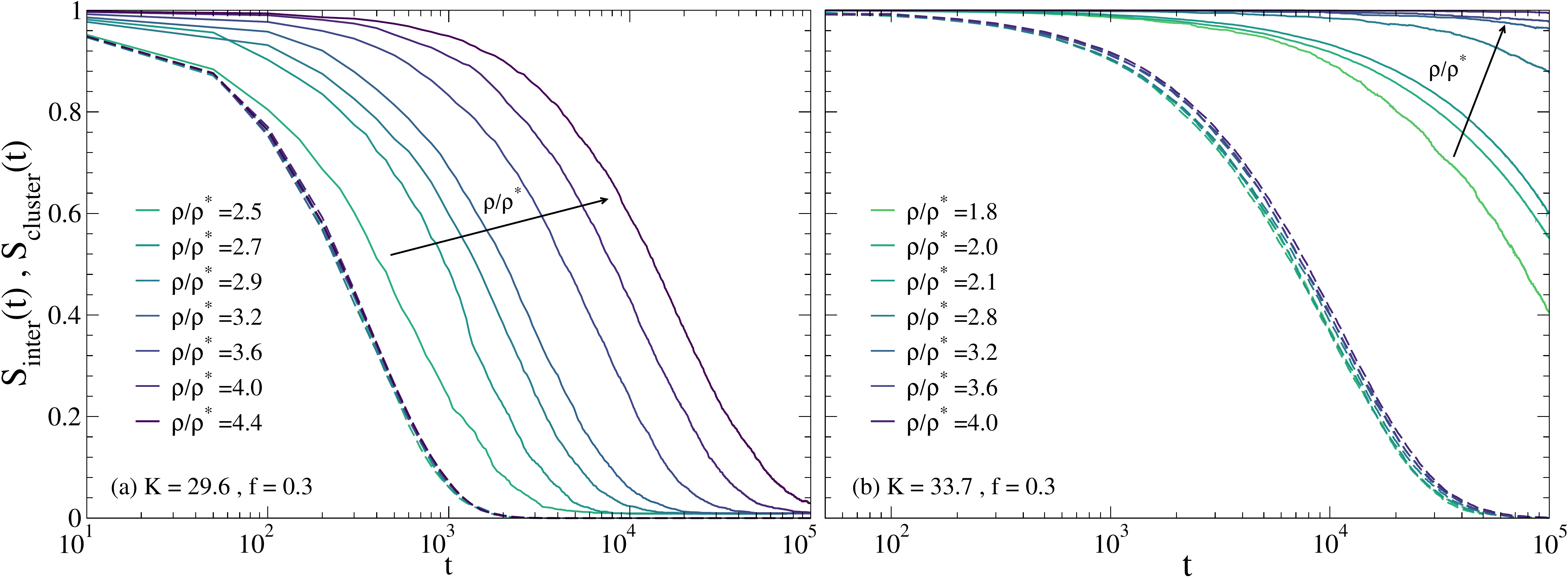}
  \caption{Intermolecular bond ($S_{\rm inter}(t)$, dashed lines) and cluster ($S_{\rm cluster}(t)$, solid lines) self-correlation function for $f=0.3$ and (a) $K=29.6$,  (b) $K=33.7$. Only densities in which a percolating cluster is at least transiently present are considered. }
  \label{fgr:bondclustercorr}
\end{figure}

Finally, we investigate the bond dynamics and the reorganization of the percolating cluster once it is formed. We first analyze whether the lifetime of intermolecular bonds is affected by gelation and whether cooperativity plays a role in the formation of intermolecular bonds. To this end, we calculate the intermolecular bond self-correlation
\begin{equation}
S_{\rm inter}(t) = \frac{\langle B_{ij}(t) B_{ij}(0)\rangle}{\langle B_{ij}(0)\rangle^2} \, ,
\end{equation}
where $B_{ij}(t) = 1$  if the reactive monomers $i$ and $j$ (belonging to different chains) form a bond at time $t$ and this has never been broken since $t=0$. If the bond has been broken at least once, then $B_{ij}(t)=0$. Furthermore, we calculate the self-correlation of the cluster 
\begin{equation}
S_{\rm cluster}(t) = \frac{\langle M_{i}(t) M_{i}(0)\rangle}{\langle M_{i}(0)\rangle^2} \, ,
\end{equation}
where $M_i(t) = 1$ if polymer $i$ has been a member of the percolating cluster at all times $0 \leq t' \leq t$ and $M_i(t) = 0$ otherwise. As such, the relaxation of $S_{\rm cluster}(t)$ provides a measure for the average dissociation time from the cluster. Figure \ref{fgr:bondclustercorr} displays the intermolecular bond self-correlation as well as the cluster self-correlation for $f=0.3$ and for the two investigated bond strengths. In both cases the intermolecular bond self-correlation is not affected by changes in the density and can be well described by an exponential decay, $S_{\rm inter}(t) \sim e^{-t/\tau}$. This exponential relaxation is also found for the intramolecular bonds (not shown), demonstrating that bond breaking is purely governed by temperature (as the bond strength is akin to an inverse temperature) and bonding is not cooperative (stretched exponentials would be otherwise observed). 

\begin{figure}[ht]
\centering
  \includegraphics[width=0.58\linewidth]{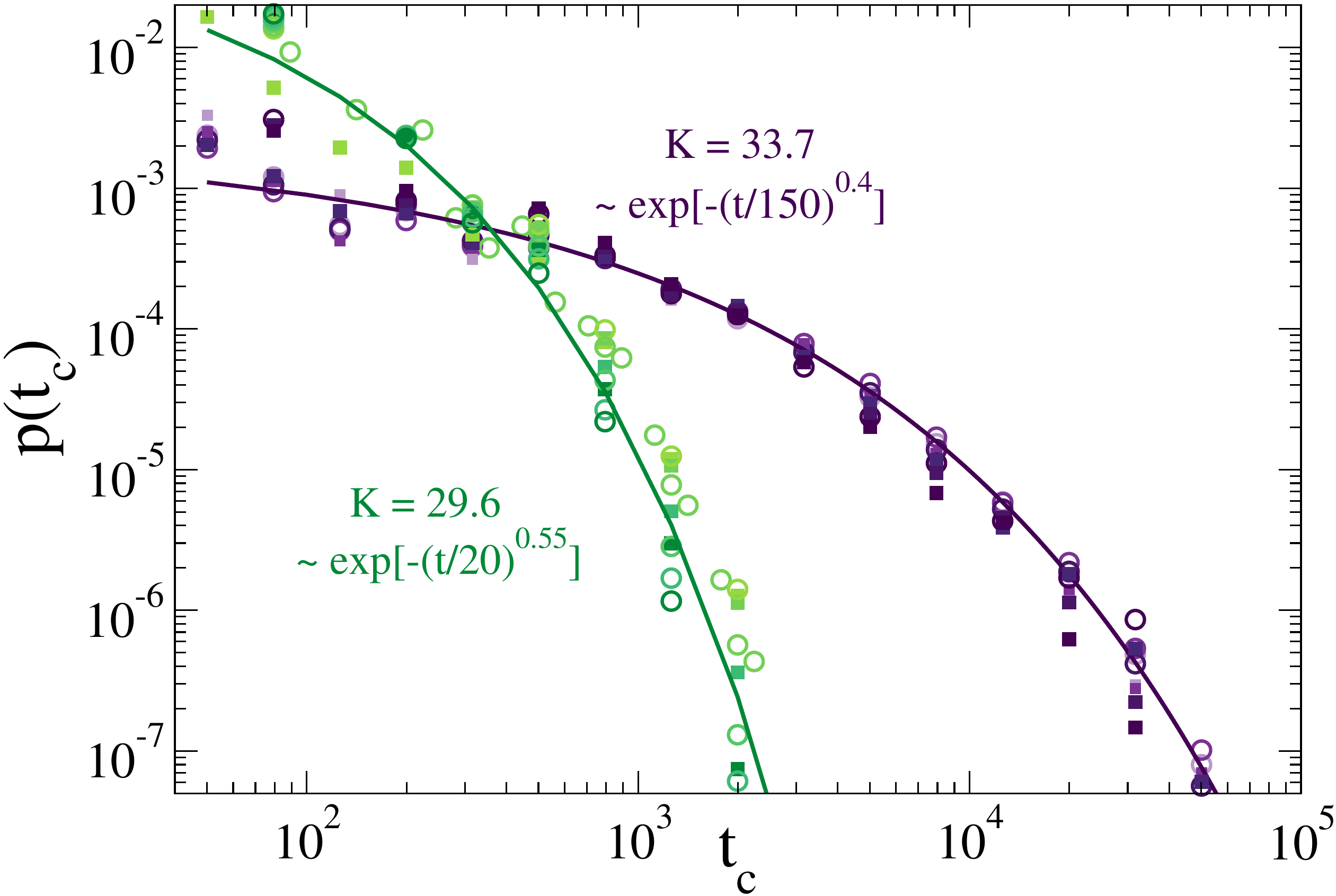}
  \caption{Distribution of times a polymer spends outside the system-spanning cluster for $K=29.6$ (green symbols) and $K=33.7$ (purple symbols). Data correspond to both $f=0.1$ (squares) and $f=0.3$ (circles) at various densities above the percolation threshold. Darker colors correspond to higher densities. Solid lines are fits to stretched exponential functions. Data are shown for the following ranges of densities: $3.2 \le \rho/\rho^\ast \le 4.4$ ($K=29.6$, $f=0.1$); $2.8 \le \rho/\rho^\ast \le 4.4$ ($K=29.6$, $f=0.3$); $1.5 \le \rho/\rho^\ast \le 3.2$ ($K=33.7$, $f=0.1$); $1.2 \le \rho/\rho^\ast \le 2.0$ ($K=33.7$, $f=0.3$). As shown in other figures, for $K=33.7$ we have performed simulations at higher concentrations. However at such concentrations the number of chains that performed excursions out of the percolating cluster was not large enough to have reasonable statistics, and the corresponding distributions $p(t_{\rm c})$ are not displayed.}
  \label{fgr:timeoutsidehist}
\end{figure}

On the other hand, the cluster self-correlation (weakly non-exponential) is strongly affected by density. For $K=33.7$, at high densities the percolating cluster becomes so stable that almost no polymers leave it within the simulation time window. This, however, does not mean that the cluster does not rearrange. Indeed, within the simulation time scale, the polymers are able to diffuse distances of more than $3R_{\rm g0} \approx 26$ even at $\rho/\rho^{\ast}=4.0$ (see Figure \ref{fgr:msd}). An average of 6 intermolecular connections per chain at this concentration 
(see Figure~\ref{fgr:intermolbonds}) allows the polymers to move through the cluster, breaking and reforming bonds, without ever detaching from it. This stability of the cluster along with its potential to rearrange could lead to an interesting behavior under external stresses, which should be investigated in future work. Furthermore, it is not clear whether intermolecular or intramolecular bonds would break first under shear, potentially leading to an interesting viscoelastic response. 

At intermediate densities, a significant number of chains (see Figure~\ref{fgr:bondclustercorr}b) detaches from the infinite cluster from time to time and moves through the system before getting reabsorbed in the cluster again. We monitor each polymer that leaves the percolating cluster at some time during the simulation, and compute the time it spends since it leaves it until it is reattached. 
The results of this analysis are shown in Figure \ref{fgr:timeoutsidehist}. Surprisingly, the distributions of times spent outside the percolating cluster are essentially independent of the  number of reactive monomers and of the density of the solution. We find a universal distribution solely dependent on the bond strength $K$ that follows a stretched exponential $p(t_{\rm c}) \sim e^{-(t/\tau_{\rm c})^\beta}$, with $\beta = 0.55$ for $K = 29.6$ and $\beta = 0.40$ for $K = 33.7$.  As shown in Figure~\ref{fgr:diffusion} the diffusivity strongly depends on 
the density, e.g., for the case $K=33.7$ and $f=0.1$ the diffusivity decays by an order of magnitude in the range $1.5 \le \rho/\rho^\ast \le 3.2$,
while in that same range $p(t_{\rm c})$ is at most marginally dependent on the concentration with no systematic trend.

As can be seen in Figure~\ref{fgr:timeoutsidehist} for $K=33.7$ most of the free motions between detachment and reattachment occur within time scales $t_{\rm c} < 10^4$ ($p(t_{\rm c})$ has decayed 2 orders of magnitude at that time), which as seen in the analysis of the MSD correspond to displacements of at most the polymer size. 
At the relaxation time $\tau_{\rm c} =150$ the MSD is still in the soft plateau regime, i.e., 
in that time scale the typical free chain is still localized in the region of its last detachment. In other words, free chains do not perform long flights until
they find an unbonded reactive group in the percolating cluster and are reattached to it.
This result suggests that the reattachment time is not controlled by the characteristic diffusivity over large scales
(which indeed is strongly affected by crowding), but rather by the time it takes to break a bond (essentially controlled by the energy scale $K\epsilon$) so that the new unbonded monomers in the free chain can explore their counterparts in the percolating cluster through conformational fluctuations. 

\section{IV. Conclusions}

We have investigated the gel formation of linear polymer chains decorated with functional groups with the ability to form reversible bonds. 
For this purpose, we have employed Langevin dynamics simulations and a bond potential that mimics reversible covalent bonds for the synthesis of reversible single-chain nanoparticles. We have studied in detail the competition between intra- and intermolecular bonding as a function of the concentration, finding that the replacement of intramolecular links by intermolecular ones prohibits a straightforward treatment of the system via Wertheim's theory and its use for elucidating the complete phase diagram. The formation of a system-spanning cluster takes place above the overlap concentration and seems to converge to it by increasing the bond strength and fraction of reactive groups. 
%The scaling of the cluster size distribution is consistent with non mean-field expectations.
 
Surprisingly, the introduction of intermolecular bonds induces a non-monotonic dependence of the radius of gyration on the density for high bond strengths, while the polymers shrink monotonically for lower bond strengths. In all the cases the variation of the polymer size with respect to high dilution conditions is at most 5\%, even at 4 times the overlap concentration. Concomitantly, the scaling properties
of the polymer conformations are weakly perturbed, and beyond the percolation transition they can be approximately described by self-avoiding random walk statistics irrespective of the bond strength. These results for the intramolecular conformations in the reversible systems are rather different from those in crowded solutions and melts of irreversible SCNPs with no intermolecular bonds.
In such systems the purely intramolecular permanent cross-links prevent concatenation of loops, leading to a collapse of purely entropic origin to crumpled globular conformations, in order to optimize packing in the solution. In the reversible case the intramolecular loops, creating topological interactions between two chains, can be broken and concatenated later with other loops
when a new intramolecular bond is formed. Under such conditions the SCNPs become more open and penetrable, and the microsegregration seen in irreversible SCNPs is circumvented, making the system significantly more compressible.  

Finally, we have demonstrated that the dynamics of the system display the soft caging regime expected for gel-forming materials. The reorganization dynamics of the percolating cluster is mostly mediated by the diffusion, through breaking and formation of bonds, of chains that, without detaching from the percolating cluster, are able to explore distances of several times their size. A remarkable universal behavior is found: the time a polymer spends out of the percolating cluster is independent of the density and solely depends on the bond lifetime, which suggests that reattachment to the percolating cluster is limited by the availability of free reactive groups and not by the diffusion of the free chain.  

In conclusion, our results present valuable preliminary insights into the gelling process of reversibly cross-linking polymers with randomly distributed functional groups. The competition between intra- and intermolecular bonds leads to complex structural rearrangements purely governed by entropical contributions. Systems in which such a competition is present have not been studied extensively in the literature to this date. We hope that our results will motivate further research efforts in this direction. 
On the other hand, the possibility of combining advanced functions (catalysis, luminescence, etc), polymer flexibility, dynamic bonds and assembly into equilibrium gels make reversible SCNPs potentially attractive systems as building blocks for the design of smart materials with self-healing properties.

\begin{acknowledgement}

We acknowledge support from the projects PGC2018-094548-B-I00 (MCIU/AEI/FEDER, UE) and IT-1175-19 (Basque Government, Spain).
M.F. acknowledges travel grants from Materials Physics Center (San Sebasti\'{a}n, Spain).

\end{acknowledgement}

%\newpage
%\bibliographystyle{ieeetr}

%\bibliography{SCNP.bib}

\begin{mcitethebibliography}{81}
\providecommand*\natexlab[1]{#1}
\providecommand*\mciteSetBstSublistMode[1]{}
\providecommand*\mciteSetBstMaxWidthForm[2]{}
\providecommand*\mciteBstWouldAddEndPuncttrue
  {\def\EndOfBibitem{\unskip.}}
\providecommand*\mciteBstWouldAddEndPunctfalse
  {\let\EndOfBibitem\relax}
\providecommand*\mciteSetBstMidEndSepPunct[3]{}
\providecommand*\mciteSetBstSublistLabelBeginEnd[3]{}
\providecommand*\EndOfBibitem{}
\mciteSetBstSublistMode{f}
\mciteSetBstMaxWidthForm{subitem}{(\alph{mcitesubitemcount})}
\mciteSetBstSublistLabelBeginEnd
  {\mcitemaxwidthsubitemform\space}
  {\relax}
  {\relax}

\bibitem[Lyon \latin{et~al.}(2015)Lyon, Prasher, Hanlon, Tuten, Tooley, Frank,
  and Berda]{Lyon2015}
Lyon,~C.~K.; Prasher,~A.; Hanlon,~A.~M.; Tuten,~B.~T.; Tooley,~C.~A.;
  Frank,~P.~G.; Berda,~E.~B. A brief user's guide to single-chain
  nanoparticles. \emph{Polymer Chemistry} \textbf{2015}, \emph{6},
  181--197\relax
\mciteBstWouldAddEndPuncttrue
\mciteSetBstMidEndSepPunct{\mcitedefaultmidpunct}
{\mcitedefaultendpunct}{\mcitedefaultseppunct}\relax
\EndOfBibitem
\bibitem[Gonzalez-Burgos \latin{et~al.}(2015)Gonzalez-Burgos, Latorre-Sanchez,
  and Pomposo]{Gonzalez-Burgos2015}
Gonzalez-Burgos,~M.; Latorre-Sanchez,~A.; Pomposo,~J.~A. Advances in single
  chain technology. \emph{Chemical Society Reviews} \textbf{2015}, \emph{44},
  6122--6142\relax
\mciteBstWouldAddEndPuncttrue
\mciteSetBstMidEndSepPunct{\mcitedefaultmidpunct}
{\mcitedefaultendpunct}{\mcitedefaultseppunct}\relax
\EndOfBibitem
\bibitem[Liu \latin{et~al.}(2015)Liu, Pauloehrl, Presolski, Albertazzi,
  Palmans, and Meijer]{Liu2015}
Liu,~Y.; Pauloehrl,~T.; Presolski,~S.~I.; Albertazzi,~L.; Palmans,~A.~R.;
  Meijer,~E. Modular synthetic platform for the construction of functional
  single-chain polymeric nanoparticles: from aqueous catalysis to
  photosensitization. \emph{Journal of the American Chemical Society}
  \textbf{2015}, \emph{137}, 13096--13105\relax
\mciteBstWouldAddEndPuncttrue
\mciteSetBstMidEndSepPunct{\mcitedefaultmidpunct}
{\mcitedefaultendpunct}{\mcitedefaultseppunct}\relax
\EndOfBibitem
\bibitem[Altintas and Barner-Kowollik(2016)Altintas, and
  Barner-Kowollik]{Altintas2016}
Altintas,~O.; Barner-Kowollik,~C. Single-chain folding of synthetic polymers: a
  critical update. \emph{Macromolecular rapid communications} \textbf{2016},
  \emph{37}, 29--46\relax
\mciteBstWouldAddEndPuncttrue
\mciteSetBstMidEndSepPunct{\mcitedefaultmidpunct}
{\mcitedefaultendpunct}{\mcitedefaultseppunct}\relax
\EndOfBibitem
\bibitem[Hanlon \latin{et~al.}(2016)Hanlon, Lyon, and Berda]{Hanlon2016}
Hanlon,~A.~M.; Lyon,~C.~K.; Berda,~E.~B. What is next in single-chain
  nanoparticles? \emph{Macromolecules} \textbf{2016}, \emph{49}, 2--14\relax
\mciteBstWouldAddEndPuncttrue
\mciteSetBstMidEndSepPunct{\mcitedefaultmidpunct}
{\mcitedefaultendpunct}{\mcitedefaultseppunct}\relax
\EndOfBibitem
\bibitem[Pomposo(2017)]{PomposoSCNPbook2017}
Pomposo,~J.~A., Ed. \emph{Single-Chain Polymer Nanoparticles: Synthesis,
  Characterization, Simulations, and Applications}; John Wiley \& Sons:
  Weinheim, Germany, 2017\relax
\mciteBstWouldAddEndPuncttrue
\mciteSetBstMidEndSepPunct{\mcitedefaultmidpunct}
{\mcitedefaultendpunct}{\mcitedefaultseppunct}\relax
\EndOfBibitem
\bibitem[Rubio-Cervilla \latin{et~al.}(2017)Rubio-Cervilla, Gonz\'{a}lez, and
  Pomposo]{RubioCervilla2017rev}
Rubio-Cervilla,~J.; Gonz\'{a}lez,~E.; Pomposo,~J.~A. Advances in Single-Chain
  Nanoparticles for Catalysis Applications. \emph{Nanomaterials} \textbf{2017},
  \emph{7}\relax
\mciteBstWouldAddEndPuncttrue
\mciteSetBstMidEndSepPunct{\mcitedefaultmidpunct}
{\mcitedefaultendpunct}{\mcitedefaultseppunct}\relax
\EndOfBibitem
\bibitem[De-La-Cuesta \latin{et~al.}(2017)De-La-Cuesta, Gonz\'{a}lez, and
  Pomposo]{DeLaCuesta2017rev}
De-La-Cuesta,~J.; Gonz\'{a}lez,~E.; Pomposo,~J.~A. Advances in Fluorescent
  Single-Chain Nanoparticles. \emph{Molecules} \textbf{2017}, \emph{22}\relax
\mciteBstWouldAddEndPuncttrue
\mciteSetBstMidEndSepPunct{\mcitedefaultmidpunct}
{\mcitedefaultendpunct}{\mcitedefaultseppunct}\relax
\EndOfBibitem
\bibitem[Pia \latin{et~al.}(2018)Pia, Kroger, and Paulusse]{Kroger2018}
Pia,~A.; Kroger,~P.; Paulusse,~J.~M. Single-chain polymer nanoparticles in
  controlled drug delivery and targeted imaging. \emph{Journal of Controlled
  Release} \textbf{2018}, \emph{286}, 326 -- 347\relax
\mciteBstWouldAddEndPuncttrue
\mciteSetBstMidEndSepPunct{\mcitedefaultmidpunct}
{\mcitedefaultendpunct}{\mcitedefaultseppunct}\relax
\EndOfBibitem
\bibitem[Rothfuss \latin{et~al.}(2018)Rothfuss, Knofel, Roesky, and
  Barner-Kowollik]{Rothfuss2018}
Rothfuss,~H.; Knofel,~N.~D.; Roesky,~P.~W.; Barner-Kowollik,~C. Single-Chain
  Nanoparticles as Catalytic Nanoreactors. \emph{Journal of the American
  Chemical Society} \textbf{2018}, \emph{140}, 5875--5881\relax
\mciteBstWouldAddEndPuncttrue
\mciteSetBstMidEndSepPunct{\mcitedefaultmidpunct}
{\mcitedefaultendpunct}{\mcitedefaultseppunct}\relax
\EndOfBibitem
\bibitem[Kr"{o}ger \latin{et~al.}(2019)Kr"{o}ger, Komil, Hamelmann, Juan,
  Stenzel, and Paulusse]{Kroger2019}
Kr"{o}ger,~A. P.~P.; Komil,~M.~I.; Hamelmann,~N.~M.; Juan,~A.; Stenzel,~M.~H.;
  Paulusse,~J. M.~J. Glucose Single-Chain Polymer Nanoparticles for Cellular
  Targeting. \emph{ACS Macro Letters} \textbf{2019}, \emph{8}, 95--101\relax
\mciteBstWouldAddEndPuncttrue
\mciteSetBstMidEndSepPunct{\mcitedefaultmidpunct}
{\mcitedefaultendpunct}{\mcitedefaultseppunct}\relax
\EndOfBibitem
\bibitem[Frisch \latin{et~al.}(2020)Frisch, Tuten, and
  Barner-Kowollik]{Frisch2020}
Frisch,~H.; Tuten,~B.~T.; Barner-Kowollik,~C. Macromolecular Superstructures: A
  Future Beyond Single Chain Nanoparticles. \emph{Israel Journal of Chemistry}
  \textbf{2020}, \emph{60}, 86--99\relax
\mciteBstWouldAddEndPuncttrue
\mciteSetBstMidEndSepPunct{\mcitedefaultmidpunct}
{\mcitedefaultendpunct}{\mcitedefaultseppunct}\relax
\EndOfBibitem
\bibitem[Verde-Sesto \latin{et~al.}(2020)Verde-Sesto, Arbe, Moreno, Cangialosi,
  Alegr\'{\i}a, Colmenero, and Pomposo]{VerdeSesto2020}
Verde-Sesto,~E.; Arbe,~A.; Moreno,~A.~J.; Cangialosi,~D.; Alegr\'{\i}a,~A.;
  Colmenero,~J.; Pomposo,~J.~A. Single-chain nanoparticles: opportunities
  provided by internal and external confinement. \emph{Mater. Horiz.}
  \textbf{2020}, \emph{7}, 2292--2313\relax
\mciteBstWouldAddEndPuncttrue
\mciteSetBstMidEndSepPunct{\mcitedefaultmidpunct}
{\mcitedefaultendpunct}{\mcitedefaultseppunct}\relax
\EndOfBibitem
\bibitem[Chen \latin{et~al.}(2020)Chen, Li, Shon, and Zimmerman]{Chen2020}
Chen,~J.; Li,~K.; Shon,~J. S.~L.; Zimmerman,~S.~C. Single-Chain Nanoparticle
  Delivers a Partner Enzyme for Concurrent and Tandem Catalysis in Cells.
  \emph{Journal of the American Chemical Society} \textbf{2020}, \emph{142},
  4565--4569\relax
\mciteBstWouldAddEndPuncttrue
\mciteSetBstMidEndSepPunct{\mcitedefaultmidpunct}
{\mcitedefaultendpunct}{\mcitedefaultseppunct}\relax
\EndOfBibitem
\bibitem[Mavila \latin{et~al.}(2016)Mavila, Eivgi, Berkovich, and
  Lemcoff]{Mavila2016}
Mavila,~S.; Eivgi,~O.; Berkovich,~I.; Lemcoff,~N.~G. Intramolecular
  cross-linking methodologies for the synthesis of polymer nanoparticles.
  \emph{Chemical reviews} \textbf{2016}, \emph{116}, 878--961\relax
\mciteBstWouldAddEndPuncttrue
\mciteSetBstMidEndSepPunct{\mcitedefaultmidpunct}
{\mcitedefaultendpunct}{\mcitedefaultseppunct}\relax
\EndOfBibitem
\bibitem[Sanchez-Sanchez and Pomposo(2014)Sanchez-Sanchez, and
  Pomposo]{Sanchez-Sanchez2014}
Sanchez-Sanchez,~A.; Pomposo,~J.~A. Single-Chain Polymer Nanoparticles via
  Non-Covalent and Dynamic Covalent Bonds. \emph{{P}art. {P}art. {S}yst.
  {C}haract.} \textbf{2014}, \emph{31}, 11--23\relax
\mciteBstWouldAddEndPuncttrue
\mciteSetBstMidEndSepPunct{\mcitedefaultmidpunct}
{\mcitedefaultendpunct}{\mcitedefaultseppunct}\relax
\EndOfBibitem
\bibitem[Artar \latin{et~al.}(2014)Artar, Huerta, Meijer, and
  Palmans]{Artar2014}
Artar,~M.; Huerta,~E.; Meijer,~E.; Palmans,~A.~R. \emph{Sequence-Controlled
  Polymers: Synthesis, Self-Assembly, and Properties}; ACS Publications, 2014;
  pp 313--325\relax
\mciteBstWouldAddEndPuncttrue
\mciteSetBstMidEndSepPunct{\mcitedefaultmidpunct}
{\mcitedefaultendpunct}{\mcitedefaultseppunct}\relax
\EndOfBibitem
\bibitem[Altintas and Barner-Kowollik(2012)Altintas, and
  Barner-Kowollik]{Altintas2012}
Altintas,~O.; Barner-Kowollik,~C. {S}ingle Chain Folding of Synthetic Polymers
  by Covalent and Non-Covalent Interactions: Current Status and Future
  Perspectives. \emph{{M}acromol. {R}apid {C}ommun.} \textbf{2012}, \emph{33},
  958--971\relax
\mciteBstWouldAddEndPuncttrue
\mciteSetBstMidEndSepPunct{\mcitedefaultmidpunct}
{\mcitedefaultendpunct}{\mcitedefaultseppunct}\relax
\EndOfBibitem
\bibitem[Seo \latin{et~al.}(2008)Seo, Beck, Paulusse, Hawker, and Kim]{Seo2008}
Seo,~M.; Beck,~B.~J.; Paulusse,~J.~M.; Hawker,~C.~J.; Kim,~S.~Y. Polymeric
  nanoparticles via noncovalent cross-linking of linear chains.
  \emph{Macromolecules} \textbf{2008}, \emph{41}, 6413--6418\relax
\mciteBstWouldAddEndPuncttrue
\mciteSetBstMidEndSepPunct{\mcitedefaultmidpunct}
{\mcitedefaultendpunct}{\mcitedefaultseppunct}\relax
\EndOfBibitem
\bibitem[Foster \latin{et~al.}(2009)Foster, Berda, and Meijer]{Foster2009}
Foster,~E.~J.; Berda,~E.~B.; Meijer,~E. Metastable supramolecular polymer
  nanoparticles via intramolecular collapse of single polymer chains.
  \emph{Journal of the American Chemical Society} \textbf{2009}, \emph{131},
  6964--6966\relax
\mciteBstWouldAddEndPuncttrue
\mciteSetBstMidEndSepPunct{\mcitedefaultmidpunct}
{\mcitedefaultendpunct}{\mcitedefaultseppunct}\relax
\EndOfBibitem
\bibitem[Terashima \latin{et~al.}(2011)Terashima, Mes, De~Greef, Gillissen,
  Besenius, Palmans, and Meijer]{Terashima2011}
Terashima,~T.; Mes,~T.; De~Greef,~T. F.~A.; Gillissen,~M. A.~J.; Besenius,~P.;
  Palmans,~A. R.~A.; Meijer,~E.~W. Single-Chain Folding of Polymers for
  Catalytic Systems in Water. \emph{Journal of the American Chemical Society}
  \textbf{2011}, \emph{133}, 4742--4745, PMID: 21405022\relax
\mciteBstWouldAddEndPuncttrue
\mciteSetBstMidEndSepPunct{\mcitedefaultmidpunct}
{\mcitedefaultendpunct}{\mcitedefaultseppunct}\relax
\EndOfBibitem
\bibitem[Hosono \latin{et~al.}(2012)Hosono, Gillissen, Li, Sheiko, Palmans, and
  Meijer]{Hosono2012}
Hosono,~N.; Gillissen,~M.~A.; Li,~Y.; Sheiko,~S.~S.; Palmans,~A.~R.; Meijer,~E.
  Orthogonal self-assembly in folding block copolymers. \emph{Journal of the
  American Chemical Society} \textbf{2012}, \emph{135}, 501--510\relax
\mciteBstWouldAddEndPuncttrue
\mciteSetBstMidEndSepPunct{\mcitedefaultmidpunct}
{\mcitedefaultendpunct}{\mcitedefaultseppunct}\relax
\EndOfBibitem
\bibitem[Burattini \latin{et~al.}(2009)Burattini, Colquhoun, Fox, Friedmann,
  Greenland, Harris, Hayes, Mackay, and Rowan]{Burattini2009}
Burattini,~S.; Colquhoun,~H.~M.; Fox,~J.~D.; Friedmann,~D.; Greenland,~B.~W.;
  Harris,~P.~J.; Hayes,~W.; Mackay,~M.~E.; Rowan,~S.~J. A self-repairing,
  supramolecular polymer system: healability as a consequence of
  donor--acceptor $\pi$--$\pi$ stacking interactions. \emph{Chemical
  communications} \textbf{2009}, 6717--6719\relax
\mciteBstWouldAddEndPuncttrue
\mciteSetBstMidEndSepPunct{\mcitedefaultmidpunct}
{\mcitedefaultendpunct}{\mcitedefaultseppunct}\relax
\EndOfBibitem
\bibitem[Appel \latin{et~al.}(2012)Appel, Dyson, del Barrio, Walsh, and
  Scherman]{Appel2012}
Appel,~E.~A.; Dyson,~J.; del Barrio,~J.; Walsh,~Z.; Scherman,~O.~A. Formation
  of single-chain polymer nanoparticles in water through host--guest
  interactions. \emph{Angewandte Chemie International Edition} \textbf{2012},
  \emph{51}, 4185--4189\relax
\mciteBstWouldAddEndPuncttrue
\mciteSetBstMidEndSepPunct{\mcitedefaultmidpunct}
{\mcitedefaultendpunct}{\mcitedefaultseppunct}\relax
\EndOfBibitem
\bibitem[Wang \latin{et~al.}(2016)Wang, Pu, and Che]{Wang2016}
Wang,~F.; Pu,~H.; Che,~X. Voltage-responsive single-chain polymer nanoparticles
  via host--guest interaction. \emph{Chemical Communications} \textbf{2016},
  \emph{52}, 3516--3519\relax
\mciteBstWouldAddEndPuncttrue
\mciteSetBstMidEndSepPunct{\mcitedefaultmidpunct}
{\mcitedefaultendpunct}{\mcitedefaultseppunct}\relax
\EndOfBibitem
\bibitem[Huh \latin{et~al.}(2006)Huh, Park, Kim, Kim, Park, and Jo]{Huh2006}
Huh,~J.; Park,~H.~J.; Kim,~K.~H.; Kim,~K.~H.; Park,~C.; Jo,~W.~H. Giant Thermal
  Tunability of the Lamellar Spacing in Block-Copolymer-Like Supramolecules
  Formed from Binary-End-Functionalized Polymer Blends. \emph{Advanced
  Materials} \textbf{2006}, \emph{18}, 624--629\relax
\mciteBstWouldAddEndPuncttrue
\mciteSetBstMidEndSepPunct{\mcitedefaultmidpunct}
{\mcitedefaultendpunct}{\mcitedefaultseppunct}\relax
\EndOfBibitem
\bibitem[Hofmeier and Schubert(2005)Hofmeier, and Schubert]{Hofmeier2005}
Hofmeier,~H.; Schubert,~U.~S. Combination of orthogonal supramolecular
  interactions in polymeric architectures. \emph{Chemical Communications}
  \textbf{2005}, 2423--2432\relax
\mciteBstWouldAddEndPuncttrue
\mciteSetBstMidEndSepPunct{\mcitedefaultmidpunct}
{\mcitedefaultendpunct}{\mcitedefaultseppunct}\relax
\EndOfBibitem
\bibitem[Gohy \latin{et~al.}(2003)Gohy, Hofmeier, Alexeev, and
  Schubert]{Gohy2003}
Gohy,~J.-F.; Hofmeier,~H.; Alexeev,~A.; Schubert,~U.~S. Aqueous micelles from
  supramolecular graft copolymers. \emph{Macromolecular Chemistry and Physics}
  \textbf{2003}, \emph{204}, 1524--1530\relax
\mciteBstWouldAddEndPuncttrue
\mciteSetBstMidEndSepPunct{\mcitedefaultmidpunct}
{\mcitedefaultendpunct}{\mcitedefaultseppunct}\relax
\EndOfBibitem
\bibitem[Tuten \latin{et~al.}(2012)Tuten, Chao, Lyon, and Berda]{Tuten2012}
Tuten,~B.~T.; Chao,~D.; Lyon,~C.~K.; Berda,~E.~B. Single-chain polymer
  nanoparticles via reversible disulfide bridges. \emph{Polymer Chemistry}
  \textbf{2012}, \emph{3}, 3068--3071\relax
\mciteBstWouldAddEndPuncttrue
\mciteSetBstMidEndSepPunct{\mcitedefaultmidpunct}
{\mcitedefaultendpunct}{\mcitedefaultseppunct}\relax
\EndOfBibitem
\bibitem[Murray and Fulton(2011)Murray, and Fulton]{Murray2011}
Murray,~B.~S.; Fulton,~D.~A. Dynamic covalent single-chain polymer
  nanoparticles. \emph{Macromolecules} \textbf{2011}, \emph{44},
  7242--7252\relax
\mciteBstWouldAddEndPuncttrue
\mciteSetBstMidEndSepPunct{\mcitedefaultmidpunct}
{\mcitedefaultendpunct}{\mcitedefaultseppunct}\relax
\EndOfBibitem
\bibitem[Sanchez-Sanchez \latin{et~al.}(2014)Sanchez-Sanchez, Fulton, and
  Pomposo]{Sanchez-Sanchez2014a}
Sanchez-Sanchez,~A.; Fulton,~D.~A.; Pomposo,~J.~A. pH-responsive single-chain
  polymer nanoparticles utilising dynamic covalent enamine bonds.
  \emph{Chemical Communications} \textbf{2014}, \emph{50}, 1871--1874\relax
\mciteBstWouldAddEndPuncttrue
\mciteSetBstMidEndSepPunct{\mcitedefaultmidpunct}
{\mcitedefaultendpunct}{\mcitedefaultseppunct}\relax
\EndOfBibitem
\bibitem[He \latin{et~al.}(2011)He, Tremblay, Lacelle, and Zhao]{He2011}
He,~J.; Tremblay,~L.; Lacelle,~S.; Zhao,~Y. Preparation of polymer single chain
  nanoparticles using intramolecular photodimerization of coumarin. \emph{Soft
  Matter} \textbf{2011}, \emph{7}, 2380--2386\relax
\mciteBstWouldAddEndPuncttrue
\mciteSetBstMidEndSepPunct{\mcitedefaultmidpunct}
{\mcitedefaultendpunct}{\mcitedefaultseppunct}\relax
\EndOfBibitem
\bibitem[Frank \latin{et~al.}(2014)Frank, Tuten, Prasher, Chao, and
  Berda]{Frank2014}
Frank,~P.~G.; Tuten,~B.~T.; Prasher,~A.; Chao,~D.; Berda,~E.~B. Intra-chain
  photodimerization of pendant anthracene units as an efficient route to
  single-chain nanoparticle fabrication. \emph{Macromolecular rapid
  communications} \textbf{2014}, \emph{35}, 249--253\relax
\mciteBstWouldAddEndPuncttrue
\mciteSetBstMidEndSepPunct{\mcitedefaultmidpunct}
{\mcitedefaultendpunct}{\mcitedefaultseppunct}\relax
\EndOfBibitem
\bibitem[Whitaker \latin{et~al.}(2013)Whitaker, Mahon, and
  Fulton]{Whitaker2013}
Whitaker,~D.~E.; Mahon,~C.~S.; Fulton,~D.~A. Thermoresponsive dynamic covalent
  single-chain polymer nanoparticles reversibly transform into a hydrogel.
  \emph{Angewandte Chemie International Edition} \textbf{2013}, \emph{52},
  956--959\relax
\mciteBstWouldAddEndPuncttrue
\mciteSetBstMidEndSepPunct{\mcitedefaultmidpunct}
{\mcitedefaultendpunct}{\mcitedefaultseppunct}\relax
\EndOfBibitem
\bibitem[Zaccarelli(2007)]{Zaccarelli2007}
Zaccarelli,~E. Colloidal gels: equilibrium and non-equilibrium routes.
  \emph{Journal of Physics: Condensed Matter} \textbf{2007}, \emph{19},
  323101\relax
\mciteBstWouldAddEndPuncttrue
\mciteSetBstMidEndSepPunct{\mcitedefaultmidpunct}
{\mcitedefaultendpunct}{\mcitedefaultseppunct}\relax
\EndOfBibitem
\bibitem[Corezzi \latin{et~al.}(2003)Corezzi, Fioretto, Puglia, and
  Kenny]{Corezzi2003}
Corezzi,~S.; Fioretto,~D.; Puglia,~D.; Kenny,~J.~M. Light scattering study of
  vitrification during the polymerization of model epoxy resins.
  \emph{Macromolecules} \textbf{2003}, \emph{36}, 5271--5278\relax
\mciteBstWouldAddEndPuncttrue
\mciteSetBstMidEndSepPunct{\mcitedefaultmidpunct}
{\mcitedefaultendpunct}{\mcitedefaultseppunct}\relax
\EndOfBibitem
\bibitem[Del~Gado \latin{et~al.}(2003)Del~Gado, Fierro, de~Arcangelis, and
  Coniglio]{DelGado2003}
Del~Gado,~E.; Fierro,~A.; de~Arcangelis,~L.; Coniglio,~A. A unifying model for
  chemical and colloidal gels. \emph{EPL (Europhysics Letters)} \textbf{2003},
  \emph{63}, 1\relax
\mciteBstWouldAddEndPuncttrue
\mciteSetBstMidEndSepPunct{\mcitedefaultmidpunct}
{\mcitedefaultendpunct}{\mcitedefaultseppunct}\relax
\EndOfBibitem
\bibitem[Saika-Voivod \latin{et~al.}(2004)Saika-Voivod, Zaccarelli, Sciortino,
  Buldyrev, and Tartaglia]{Saika-Voivod2004}
Saika-Voivod,~I.; Zaccarelli,~E.; Sciortino,~F.; Buldyrev,~S.~V.; Tartaglia,~P.
  Effect of bond lifetime on the dynamics of a short-range attractive colloidal
  system. \emph{Physical Review E} \textbf{2004}, \emph{70}, 041401\relax
\mciteBstWouldAddEndPuncttrue
\mciteSetBstMidEndSepPunct{\mcitedefaultmidpunct}
{\mcitedefaultendpunct}{\mcitedefaultseppunct}\relax
\EndOfBibitem
\bibitem[Rovigatti and Sciortino(2011)Rovigatti, and Sciortino]{Rovigatti2011}
Rovigatti,~L.; Sciortino,~F. Self and collective correlation functions in a gel
  of tetrahedral patchy particles. \emph{Molecular Physics} \textbf{2011},
  \emph{109}, 2889--2896\relax
\mciteBstWouldAddEndPuncttrue
\mciteSetBstMidEndSepPunct{\mcitedefaultmidpunct}
{\mcitedefaultendpunct}{\mcitedefaultseppunct}\relax
\EndOfBibitem
\bibitem[Groenewold and Kegel(2004)Groenewold, and Kegel]{Groenewold2004}
Groenewold,~J.; Kegel,~W. Colloidal cluster phases, gelation and nuclear
  matter. \emph{Journal of Physics: Condensed Matter} \textbf{2004}, \emph{16},
  S4877\relax
\mciteBstWouldAddEndPuncttrue
\mciteSetBstMidEndSepPunct{\mcitedefaultmidpunct}
{\mcitedefaultendpunct}{\mcitedefaultseppunct}\relax
\EndOfBibitem
\bibitem[Sciortino \latin{et~al.}(2005)Sciortino, Tartaglia, and
  Zaccarelli]{Sciortino2005}
Sciortino,~F.; Tartaglia,~P.; Zaccarelli,~E. One-dimensional cluster growth and
  branching gels in colloidal systems with short-range depletion attraction and
  screened electrostatic repulsion. \emph{The Journal of Physical Chemistry B}
  \textbf{2005}, \emph{109}, 21942--21953\relax
\mciteBstWouldAddEndPuncttrue
\mciteSetBstMidEndSepPunct{\mcitedefaultmidpunct}
{\mcitedefaultendpunct}{\mcitedefaultseppunct}\relax
\EndOfBibitem
\bibitem[Ruiz-Franco and Zaccarelli(2021)Ruiz-Franco, and Zaccarelli]{Ruiz2020}
Ruiz-Franco,~J.; Zaccarelli,~E. On the Role of Competing Interactions in
  Charged Colloids with Short-Range Attraction. \emph{Annual Review of
  Condensed Matter Physics} \textbf{2021}, \emph{12}, in advance\relax
\mciteBstWouldAddEndPuncttrue
\mciteSetBstMidEndSepPunct{\mcitedefaultmidpunct}
{\mcitedefaultendpunct}{\mcitedefaultseppunct}\relax
\EndOfBibitem
\bibitem[Zaccarelli \latin{et~al.}(2005)Zaccarelli, Buldyrev, La~Nave, Moreno,
  Saika-Voivod, Sciortino, and Tartaglia]{Zaccarelli2005}
Zaccarelli,~E.; Buldyrev,~S.; La~Nave,~E.; Moreno,~A.; Saika-Voivod,~I.;
  Sciortino,~F.; Tartaglia,~P. Model for reversible colloidal gelation.
  \emph{Physical Review Letters} \textbf{2005}, \emph{94}, 218301\relax
\mciteBstWouldAddEndPuncttrue
\mciteSetBstMidEndSepPunct{\mcitedefaultmidpunct}
{\mcitedefaultendpunct}{\mcitedefaultseppunct}\relax
\EndOfBibitem
\bibitem[Segre \latin{et~al.}(2001)Segre, Prasad, Schofield, and
  Weitz]{Segre2001}
Segre,~P.; Prasad,~V.; Schofield,~A.; Weitz,~D. Glasslike kinetic arrest at the
  colloidal-gelation transition. \emph{Physical Review Letters} \textbf{2001},
  \emph{86}, 6042\relax
\mciteBstWouldAddEndPuncttrue
\mciteSetBstMidEndSepPunct{\mcitedefaultmidpunct}
{\mcitedefaultendpunct}{\mcitedefaultseppunct}\relax
\EndOfBibitem
\bibitem[Manoharan \latin{et~al.}(2003)Manoharan, Elsesser, and
  Pine]{Manoharan2003}
Manoharan,~V.~N.; Elsesser,~M.~T.; Pine,~D.~J. Dense packing and symmetry in
  small clusters of microspheres. \emph{Science} \textbf{2003}, \emph{301},
  483--487\relax
\mciteBstWouldAddEndPuncttrue
\mciteSetBstMidEndSepPunct{\mcitedefaultmidpunct}
{\mcitedefaultendpunct}{\mcitedefaultseppunct}\relax
\EndOfBibitem
\bibitem[Cho \latin{et~al.}(2005)Cho, Yi, Lim, Kim, Manoharan, Pine, and
  Yang]{Cho2005}
Cho,~Y.-S.; Yi,~G.-R.; Lim,~J.-M.; Kim,~S.-H.; Manoharan,~V.~N.; Pine,~D.~J.;
  Yang,~S.-M. Self-organization of bidisperse colloids in water droplets.
  \emph{Journal of the American Chemical Society} \textbf{2005}, \emph{127},
  15968--15975\relax
\mciteBstWouldAddEndPuncttrue
\mciteSetBstMidEndSepPunct{\mcitedefaultmidpunct}
{\mcitedefaultendpunct}{\mcitedefaultseppunct}\relax
\EndOfBibitem
\bibitem[Biffi \latin{et~al.}(2013)Biffi, Cerbino, Bomboi, Paraboschi, Asselta,
  Sciortino, and Bellini]{Biffi2013}
Biffi,~S.; Cerbino,~R.; Bomboi,~F.; Paraboschi,~E.~M.; Asselta,~R.;
  Sciortino,~F.; Bellini,~T. Phase behavior and critical activated dynamics of
  limited-valence DNA nanostars. \emph{Proceedings of the National Academy of
  Sciences} \textbf{2013}, \emph{110}, 15633--15637\relax
\mciteBstWouldAddEndPuncttrue
\mciteSetBstMidEndSepPunct{\mcitedefaultmidpunct}
{\mcitedefaultendpunct}{\mcitedefaultseppunct}\relax
\EndOfBibitem
\bibitem[Rovigatti \latin{et~al.}(2014)Rovigatti, Smallenburg, Romano, and
  Sciortino]{Rovigatti2014}
Rovigatti,~L.; Smallenburg,~F.; Romano,~F.; Sciortino,~F. Gels of DNA nanostars
  never crystallize. \emph{ACS nano} \textbf{2014}, \emph{8}, 3567--3574\relax
\mciteBstWouldAddEndPuncttrue
\mciteSetBstMidEndSepPunct{\mcitedefaultmidpunct}
{\mcitedefaultendpunct}{\mcitedefaultseppunct}\relax
\EndOfBibitem
\bibitem[Wertheim(1984)]{Wertheim1984}
Wertheim,~M. Fluids with highly directional attractive forces. I. Statistical
  thermodynamics. \emph{Journal of statistical physics} \textbf{1984},
  \emph{35}, 19--34\relax
\mciteBstWouldAddEndPuncttrue
\mciteSetBstMidEndSepPunct{\mcitedefaultmidpunct}
{\mcitedefaultendpunct}{\mcitedefaultseppunct}\relax
\EndOfBibitem
\bibitem[Wertheim(1984)]{Wertheim1984a}
Wertheim,~M. Fluids with highly directional attractive forces. II.
  Thermodynamic perturbation theory and integral equations. \emph{Journal of
  statistical physics} \textbf{1984}, \emph{35}, 35--47\relax
\mciteBstWouldAddEndPuncttrue
\mciteSetBstMidEndSepPunct{\mcitedefaultmidpunct}
{\mcitedefaultendpunct}{\mcitedefaultseppunct}\relax
\EndOfBibitem
\bibitem[Bianchi \latin{et~al.}(2006)Bianchi, Largo, Tartaglia, Zaccarelli, and
  Sciortino]{Bianchi2006}
Bianchi,~E.; Largo,~J.; Tartaglia,~P.; Zaccarelli,~E.; Sciortino,~F. Phase
  diagram of patchy colloids: Towards empty liquids. \emph{Physical Review
  Letters} \textbf{2006}, \emph{97}, 168301\relax
\mciteBstWouldAddEndPuncttrue
\mciteSetBstMidEndSepPunct{\mcitedefaultmidpunct}
{\mcitedefaultendpunct}{\mcitedefaultseppunct}\relax
\EndOfBibitem
\bibitem[Whitesides and Boncheva(2002)Whitesides, and Boncheva]{Whitesides2002}
Whitesides,~G.~M.; Boncheva,~M. Beyond molecules: Self-assembly of mesoscopic
  and macroscopic components. \emph{Proceedings of the National Academy of
  Sciences} \textbf{2002}, \emph{99}, 4769--4774\relax
\mciteBstWouldAddEndPuncttrue
\mciteSetBstMidEndSepPunct{\mcitedefaultmidpunct}
{\mcitedefaultendpunct}{\mcitedefaultseppunct}\relax
\EndOfBibitem
\bibitem[Glotzer(2004)]{Glotzer2004}
Glotzer,~S.~C. Some Assembly Required. \emph{Science} \textbf{2004},
  \emph{306}, 419--420\relax
\mciteBstWouldAddEndPuncttrue
\mciteSetBstMidEndSepPunct{\mcitedefaultmidpunct}
{\mcitedefaultendpunct}{\mcitedefaultseppunct}\relax
\EndOfBibitem
\bibitem[Moreno \latin{et~al.}(2016)Moreno, Lo~Verso, Arbe, Pomposo, and
  Colmenero]{Moreno2016}
Moreno,~A.~J.; Lo~Verso,~F.; Arbe,~A.; Pomposo,~J.~A.; Colmenero,~J.
  Concentrated Solutions of Single-Chain Nanoparticles: A Simple Model for
  Intrinsically Disordered Proteins under Crowding Conditions. \emph{{J}.
  {P}hys. {C}hem. {L}ett.} \textbf{2016}, \emph{7}, 838--844\relax
\mciteBstWouldAddEndPuncttrue
\mciteSetBstMidEndSepPunct{\mcitedefaultmidpunct}
{\mcitedefaultendpunct}{\mcitedefaultseppunct}\relax
\EndOfBibitem
\bibitem[Kremer and Grest(1990)Kremer, and Grest]{Kremer1990}
Kremer,~K.; Grest,~G.~S. Dynamics of entangled linear polymer melts: A
  molecular-dynamics simulation. \emph{{J}. {C}hem. {P}hys.} \textbf{1990},
  \emph{92}, 5057--5086\relax
\mciteBstWouldAddEndPuncttrue
\mciteSetBstMidEndSepPunct{\mcitedefaultmidpunct}
{\mcitedefaultendpunct}{\mcitedefaultseppunct}\relax
\EndOfBibitem
\bibitem[Rubinstein and Colby(2003)Rubinstein, and Colby]{Rubinstein2003}
Rubinstein,~M.; Colby,~R.~H. \emph{{P}olymer {P}hysics}; {O}xford {U}niversity
  {P}ress: {O}xford, {U}.{K}., 2003; Vol.~23\relax
\mciteBstWouldAddEndPuncttrue
\mciteSetBstMidEndSepPunct{\mcitedefaultmidpunct}
{\mcitedefaultendpunct}{\mcitedefaultseppunct}\relax
\EndOfBibitem
\bibitem[Moreno \latin{et~al.}(2013)Moreno, Lo~Verso, Sanchez-Sanchez, Arbe,
  Colmenero, and Pomposo]{Moreno2013}
Moreno,~A.~J.; Lo~Verso,~F.; Sanchez-Sanchez,~A.; Arbe,~A.; Colmenero,~J.;
  Pomposo,~J.~A. Advantages of Orthogonal Folding of Single Polymer Chains to
  Soft Nanoparticles. \emph{{M}acromolecules} \textbf{2013}, \emph{46},
  9748--9759\relax
\mciteBstWouldAddEndPuncttrue
\mciteSetBstMidEndSepPunct{\mcitedefaultmidpunct}
{\mcitedefaultendpunct}{\mcitedefaultseppunct}\relax
\EndOfBibitem
\bibitem[Formanek and Moreno(2017)Formanek, and Moreno]{Formanek2017}
Formanek,~M.; Moreno,~A.~J. Effects of precursor topology and synthesis under
  crowding conditions on the structure of single-chain polymer nanoparticles.
  \emph{Soft matter} \textbf{2017}, \emph{13}, 6430--6438\relax
\mciteBstWouldAddEndPuncttrue
\mciteSetBstMidEndSepPunct{\mcitedefaultmidpunct}
{\mcitedefaultendpunct}{\mcitedefaultseppunct}\relax
\EndOfBibitem
\bibitem[Formanek and Moreno(2019)Formanek, and Moreno]{Formanek2019}
Formanek,~M.; Moreno,~A.~J. Single-Chain Nanoparticles under Homogeneous Shear
  Flow. \emph{Macromolecules} \textbf{2019}, \emph{52}, 1821--1831\relax
\mciteBstWouldAddEndPuncttrue
\mciteSetBstMidEndSepPunct{\mcitedefaultmidpunct}
{\mcitedefaultendpunct}{\mcitedefaultseppunct}\relax
\EndOfBibitem
\bibitem[Formanek and Moreno(2019)Formanek, and Moreno]{Formanek2019a}
Formanek,~M.; Moreno,~A.~J. Crowded Solutions of Single-Chain Nanoparticles
  under Shear Flow. \emph{arXiv preprint arXiv:1910.01425} \textbf{2019},
  \relax
\mciteBstWouldAddEndPunctfalse
\mciteSetBstMidEndSepPunct{\mcitedefaultmidpunct}
{}{\mcitedefaultseppunct}\relax
\EndOfBibitem
\bibitem[Smallenburg and Sciortino(2013)Smallenburg, and
  Sciortino]{Smallenburg2013}
Smallenburg,~F.; Sciortino,~F. Liquids more stable than crystals in particles
  with limited valence and flexible bonds. \emph{Nature Physics} \textbf{2013},
  \emph{9}, 554\relax
\mciteBstWouldAddEndPuncttrue
\mciteSetBstMidEndSepPunct{\mcitedefaultmidpunct}
{\mcitedefaultendpunct}{\mcitedefaultseppunct}\relax
\EndOfBibitem
\bibitem[Locatelli \latin{et~al.}(2017)Locatelli, Handle, Likos, Sciortino, and
  Rovigatti]{Locatelli2017}
Locatelli,~E.; Handle,~P.~H.; Likos,~C.~N.; Sciortino,~F.; Rovigatti,~L.
  Condensation and Demixing in Solutions of DNA Nanostars and Their Mixtures.
  \emph{ACS Nano} \textbf{2017}, \emph{11}, 2094--2102, PMID: 28157331\relax
\mciteBstWouldAddEndPuncttrue
\mciteSetBstMidEndSepPunct{\mcitedefaultmidpunct}
{\mcitedefaultendpunct}{\mcitedefaultseppunct}\relax
\EndOfBibitem
\bibitem[Izaguirre \latin{et~al.}(2001)Izaguirre, Catarello, Wozniak, and
  Skeel]{Izaguirre2001}
Izaguirre,~J.~A.; Catarello,~D.~P.; Wozniak,~J.~M.; Skeel,~R.~D. Langevin
  stabilization of molecular dynamics. \emph{J. Chem. Phys.} \textbf{2001},
  \emph{114}, 2090--2098\relax
\mciteBstWouldAddEndPuncttrue
\mciteSetBstMidEndSepPunct{\mcitedefaultmidpunct}
{\mcitedefaultendpunct}{\mcitedefaultseppunct}\relax
\EndOfBibitem
\bibitem[Gonzalez-Burgos \latin{et~al.}(2018)Gonzalez-Burgos, Arbe, Moreno,
  Pomposo, Radulescu, and Colmenero]{Gonzalez-Burgos2018}
Gonzalez-Burgos,~M.; Arbe,~A.; Moreno,~A.~J.; Pomposo,~J.~A.; Radulescu,~A.;
  Colmenero,~J. Crowding the Environment of Single-Chain Nanoparticles: A
  Combined Study by SANS and Simulations. \emph{Macromolecules} \textbf{2018},
  \emph{51}, 1573--1585\relax
\mciteBstWouldAddEndPuncttrue
\mciteSetBstMidEndSepPunct{\mcitedefaultmidpunct}
{\mcitedefaultendpunct}{\mcitedefaultseppunct}\relax
\EndOfBibitem
\bibitem[Sukumaran \latin{et~al.}(2005)Sukumaran, Grest, Kremer, and
  Everaers]{Sukumaran2005}
Sukumaran,~S.~K.; Grest,~G.~S.; Kremer,~K.; Everaers,~R. Identifying the
  primitive path mesh in entangled polymer liquids. \emph{Journal of Polymer
  Science Part B: Polymer Physics} \textbf{2005}, \emph{43}, 917--933\relax
\mciteBstWouldAddEndPuncttrue
\mciteSetBstMidEndSepPunct{\mcitedefaultmidpunct}
{\mcitedefaultendpunct}{\mcitedefaultseppunct}\relax
\EndOfBibitem
\bibitem[Rold{\'a}n-Vargas \latin{et~al.}(2013)Rold{\'a}n-Vargas, Smallenburg,
  Kob, and Sciortino]{Roldan-Vargas2013}
Rold{\'a}n-Vargas,~S.; Smallenburg,~F.; Kob,~W.; Sciortino,~F. Phase diagram of
  a reentrant gel of patchy particles. \emph{The Journal of chemical physics}
  \textbf{2013}, \emph{139}, 244910\relax
\mciteBstWouldAddEndPuncttrue
\mciteSetBstMidEndSepPunct{\mcitedefaultmidpunct}
{\mcitedefaultendpunct}{\mcitedefaultseppunct}\relax
\EndOfBibitem
\bibitem[Marshall \latin{et~al.}(2012)Marshall, Ballal, and
  Chapman]{Marshall2012}
Marshall,~B.~D.; Ballal,~D.; Chapman,~W.~G. Wertheim's association theory
  applied to one site patchy colloids: Beyond the single bonding condition.
  \emph{The Journal of chemical physics} \textbf{2012}, \emph{137},
  104909\relax
\mciteBstWouldAddEndPuncttrue
\mciteSetBstMidEndSepPunct{\mcitedefaultmidpunct}
{\mcitedefaultendpunct}{\mcitedefaultseppunct}\relax
\EndOfBibitem
\bibitem[Sciortino(2019)]{Sciortino2019}
Sciortino,~F. Entropy in Self-assembly. \emph{Rivista del Nuovo Cimento}
  \textbf{2019}, \emph{42}, 511--548\relax
\mciteBstWouldAddEndPuncttrue
\mciteSetBstMidEndSepPunct{\mcitedefaultmidpunct}
{\mcitedefaultendpunct}{\mcitedefaultseppunct}\relax
\EndOfBibitem
\bibitem[Sciortino \latin{et~al.}(2020)Sciortino, Zhang, Gang, and
  Kumar]{Sciortino2020}
Sciortino,~F.; Zhang,~Y.; Gang,~O.; Kumar,~S.~K. Combinatorial-Entropy-Driven
  Aggregation in DNA-Grafted Nanoparticles. \emph{ACS Nano} \textbf{2020},
  \emph{14}, 5628--5635\relax
\mciteBstWouldAddEndPuncttrue
\mciteSetBstMidEndSepPunct{\mcitedefaultmidpunct}
{\mcitedefaultendpunct}{\mcitedefaultseppunct}\relax
\EndOfBibitem
\bibitem[Flory(1941)]{Flory1941}
Flory,~P. Molecular size in three dimensional polymers. I. Gelation. II.
  Tri-functional branching units. III. Tetrafunctional branching units.
  \emph{J. Am. Chem. Soc.} \textbf{1941}, \emph{63}, 3083\relax
\mciteBstWouldAddEndPuncttrue
\mciteSetBstMidEndSepPunct{\mcitedefaultmidpunct}
{\mcitedefaultendpunct}{\mcitedefaultseppunct}\relax
\EndOfBibitem
\bibitem[Stockmayer(1943)]{Stockmayer1943}
Stockmayer,~W.~H. Theory of molecular size distribution and gel formation in
  branched-chain polymers. \emph{The Journal of chemical physics}
  \textbf{1943}, \emph{11}, 45--55\relax
\mciteBstWouldAddEndPuncttrue
\mciteSetBstMidEndSepPunct{\mcitedefaultmidpunct}
{\mcitedefaultendpunct}{\mcitedefaultseppunct}\relax
\EndOfBibitem
\bibitem[Stockmayer(1944)]{Stockmayer1944}
Stockmayer,~W.~H. Theory of molecular size distribution and gel formation in
  branched polymers II. General cross linking. \emph{The Journal of Chemical
  Physics} \textbf{1944}, \emph{12}, 125--131\relax
\mciteBstWouldAddEndPuncttrue
\mciteSetBstMidEndSepPunct{\mcitedefaultmidpunct}
{\mcitedefaultendpunct}{\mcitedefaultseppunct}\relax
\EndOfBibitem
\bibitem[Stauffer and Aharony(1992)Stauffer, and Aharony]{Stauffer1992}
Stauffer,~D.; Aharony,~A. \emph{Introduction to percolation theory}; Taylor \&
  Francis, London, 1992\relax
\mciteBstWouldAddEndPuncttrue
\mciteSetBstMidEndSepPunct{\mcitedefaultmidpunct}
{\mcitedefaultendpunct}{\mcitedefaultseppunct}\relax
\EndOfBibitem
\bibitem[Stauffer(1975)]{Stauffer1975}
Stauffer,~D. Violation of dynamical scaling for randomly dilute Ising
  ferromagnets near percolation threshold. \emph{Physical Review Letters}
  \textbf{1975}, \emph{35}, 394\relax
\mciteBstWouldAddEndPuncttrue
\mciteSetBstMidEndSepPunct{\mcitedefaultmidpunct}
{\mcitedefaultendpunct}{\mcitedefaultseppunct}\relax
\EndOfBibitem
\bibitem[Kirkpatrick(1976)]{Kirkpatrick1976}
Kirkpatrick,~S. Percolation phenomena in higher dimensions: Approach to the
  mean-field limit. \emph{Physical Review Letters} \textbf{1976}, \emph{36},
  69\relax
\mciteBstWouldAddEndPuncttrue
\mciteSetBstMidEndSepPunct{\mcitedefaultmidpunct}
{\mcitedefaultendpunct}{\mcitedefaultseppunct}\relax
\EndOfBibitem
\bibitem[Stauffer(1979)]{Stauffer1979}
Stauffer,~D. Scaling theory of percolation clusters. \emph{Physics reports}
  \textbf{1979}, \emph{54}, 1--74\relax
\mciteBstWouldAddEndPuncttrue
\mciteSetBstMidEndSepPunct{\mcitedefaultmidpunct}
{\mcitedefaultendpunct}{\mcitedefaultseppunct}\relax
\EndOfBibitem
\bibitem[Halverson \latin{et~al.}(2011)Halverson, Lee, Grest, Grosberg, and
  Kremer]{Halverson2011}
Halverson,~J.~D.; Lee,~W.~B.; Grest,~G.~S.; Grosberg,~A.~Y.; Kremer,~K.
  Molecular dynamics simulation study of nonconcatenated ring polymers in a
  melt. I. Statics. \emph{{J}. {C}hem. {P}hys.} \textbf{2011}, \emph{134},
  204904\relax
\mciteBstWouldAddEndPuncttrue
\mciteSetBstMidEndSepPunct{\mcitedefaultmidpunct}
{\mcitedefaultendpunct}{\mcitedefaultseppunct}\relax
\EndOfBibitem
\bibitem[Pomposo \latin{et~al.}(2017)Pomposo, Rubio-Cervilla, Moreno, Lo~Verso,
  Bacova, Arbe, and Colmenero]{Pomposo2017}
Pomposo,~J.~A.; Rubio-Cervilla,~J.; Moreno,~A.~J.; Lo~Verso,~F.; Bacova,~P.;
  Arbe,~A.; Colmenero,~J. Folding Single Chains to Single-Chain Nanoparticles
  via Reversible Interactions: What Size Reduction Can One Expect?
  \emph{Macromolecules} \textbf{2017}, \emph{50}, 1732--1739\relax
\mciteBstWouldAddEndPuncttrue
\mciteSetBstMidEndSepPunct{\mcitedefaultmidpunct}
{\mcitedefaultendpunct}{\mcitedefaultseppunct}\relax
\EndOfBibitem
\bibitem[Barrat and Hansen(2003)Barrat, and Hansen]{BarratHansen2003}
Barrat,~J.~L.; Hansen,~J.~P. \emph{Basic Concepts for Simple and Complex
  Liquids}; Cambridge University Press: Cambridge, U.K., 2003\relax
\mciteBstWouldAddEndPuncttrue
\mciteSetBstMidEndSepPunct{\mcitedefaultmidpunct}
{\mcitedefaultendpunct}{\mcitedefaultseppunct}\relax
\EndOfBibitem
\bibitem[not()]{noterouse}
One should note that the absence, prior to the plateau, of an intermediate
  Rouse-like regime (${\rm MSD }\sim t^{0.5}$) is due to the use of a low
  friction constant in order to speed up equilibration and sampling.\relax
\mciteBstWouldAddEndPunctfalse
\mciteSetBstMidEndSepPunct{\mcitedefaultmidpunct}
{}{\mcitedefaultseppunct}\relax
\EndOfBibitem
\end{mcitethebibliography}

\providecommand{\latin}[1]{#1}
\providecommand*\mcitethebibliography{\thebibliography}
\csname @ifundefined\endcsname{endmcitethebibliography}
  {\let\endmcitethebibliography\endthebibliography}{}

\newpage

\begin{center}
\LARGE
{\bf Graphical TOC Entry}
\end{center}
\vspace{2 cm}

\begin{figure}[ht]
\centering
  \includegraphics[width=17.6cm,height=8.3cm]{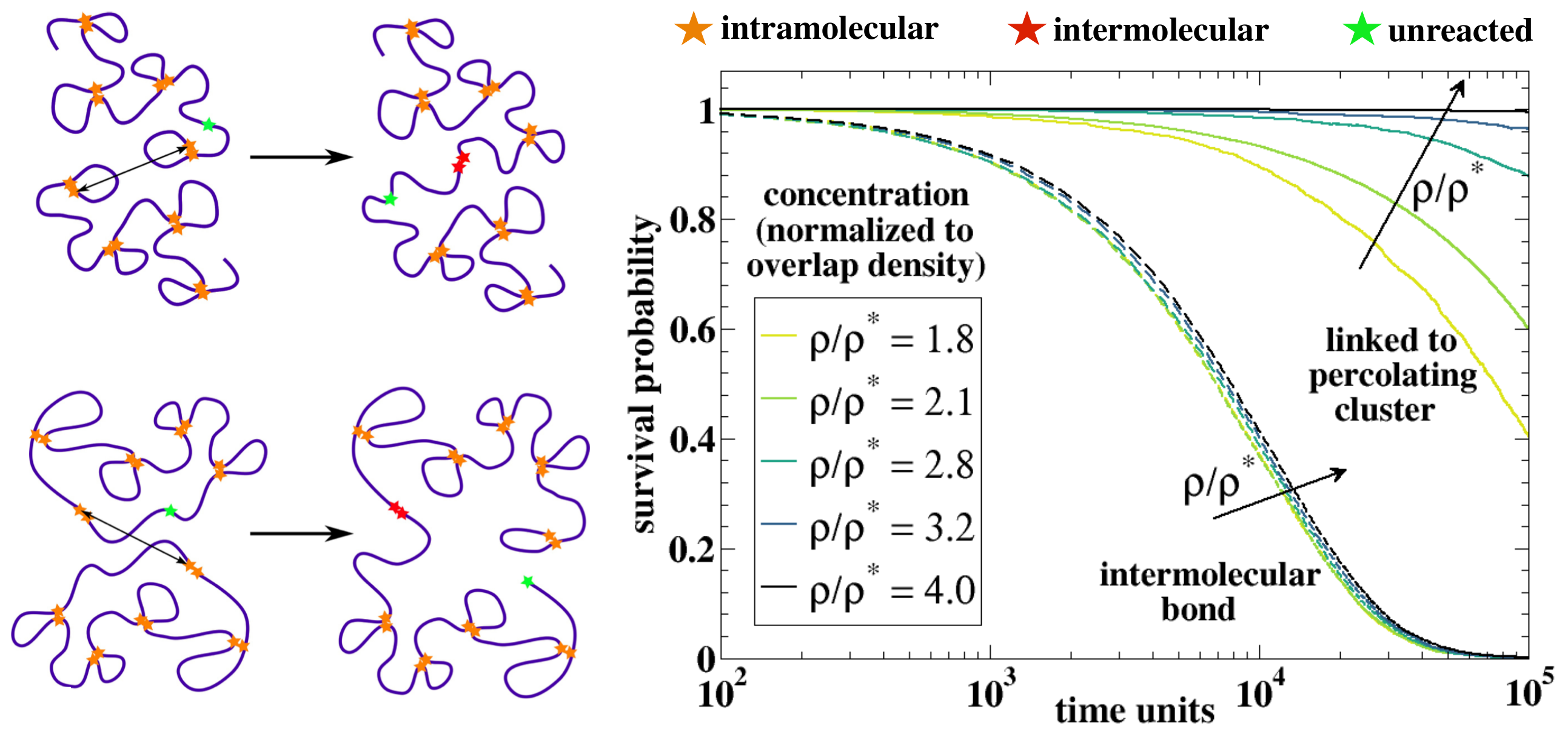}
\end{figure}

\end{document}